\documentclass[reprint,superscriptaddress,amssymb,aps, pra, twocolumn,graphicx]{revtex4-1}

\usepackage{graphicx}% Include figure files
\usepackage{dcolumn}% Align table columns on decimal point
\usepackage{bm}% bold math
\usepackage{hyperref}

\usepackage{amsmath} 
\usepackage{mathrsfs}
\usepackage[cmintegrals]{newtxmath}
\usepackage[mathlines]{lineno}

\begin{document}

%\bibliographystyle{unsrt}
%\preprint{APS/123-QED}

\title{Dynamical encircling exceptional point in largely detuned multimode optomechanical system}

\author{Dan Long}
\email{These authors contributed equally to this work}
\affiliation{Department of Physics, State Key Laboratory of Low-Dimensional Quantum Physics, Tsinghua University, Beijing 100084, China}

\author{Xuan Mao}
\email{These authors contributed equally to this work}
\affiliation{Department of Physics, State Key Laboratory of Low-Dimensional Quantum Physics, Tsinghua University, Beijing 100084, China}

\author{Guo-Qing Qin}
\affiliation{Department of Physics, State Key Laboratory of Low-Dimensional Quantum Physics, Tsinghua University, Beijing 100084, China}

\author{Hao Zhang}
\affiliation{Purple Mountain Laboratories, Nanjing 211111, China}

\author{Min Wang}
\affiliation{Beijing Academy of Quantum Information Sciences, Beijing 100193, China}

\author{Gui-Qin Li}
\affiliation{Department of Physics, State Key Laboratory of Low-Dimensional Quantum Physics, Tsinghua University, Beijing 100084, China}
\affiliation{Frontier Science Center for Quantum Information, Beijing 100084, China}

\author{Gui-Lu Long}
\email{gllong@tsinghua.edu.cn}
\affiliation{Department of Physics, State Key Laboratory of Low-Dimensional Quantum Physics, Tsinghua University, Beijing 100084, China}
\affiliation{Beijing Academy of Quantum Information Sciences, Beijing 100193, China}
\affiliation{Frontier Science Center for Quantum Information, Beijing 100084, China}
\affiliation{Beijing National Research Center for Information Science and Technology, Beijing 100084, China}
\affiliation{School of Information, Tsinghua University, Beijing 100084, China}

\date{\today}

\begin{abstract}

Dynamical encircling exceptional point(EP) shows a number of intriguing physical phenomena and its potential applications. To enrich the manipulations of optical systems in experiment, here, we study the dynamical encircling EP, i.e. state transfer process, in largely detuned multimode optomechanical system. The process of state transfer has been investigated with different factors about the location of start point, the orientation and the initial state of the trajectories around the EP in parameter space. Results show that the nonreciprocal and the chiral topological energy transfer between two optical modes are performed successfully by tuning the effective optomechanical coupling in the multimode system with large detuning. Moreover, the factor of evolution speed about system parameters is also discussed. Our work demonstrates the fundamental physics around EP in large detuning domain of multimode optomechanical system and provides an alternative for manipulating of optical modes in non-hermitian system.

\end{abstract}


\maketitle

%\tableofcontents

\section{INTRODUCTION \label{introduction}}

Exceptional points (EPs)\cite{RN2.1.1,RN1.13.1}, non-Hermitian degeneracy points at which eigenvalues and eigenvectors coalesce simultaneously\cite{RN1.1}, have been promoted in recent years\cite{RN1.2.1,RN1.2.2}. Significantly different from diabolic points (DPs) whose eigenvalues coalesce while the associated eigenvectors can always be chosen to be orthogonal in Hermitian system\cite{RN1.3.1}, plenty of practical applications\cite{RN3.1.5,RN1.4.1,RN1.4.2,RN1.4.3,RN1.2.2,RN1.5.1,RN1.5.3,RN1.5.2,RN1.5.4,RN1.5.5} have been utilized such as phonon laser\cite{RN1.4.1,RN1.4.2,RN1.4.3}, and ultra-sensitive sensors based on EP\cite{RN1.2.2,RN1.5.1,RN1.5.3,RN1.5.2,RN1.5.4,RN1.5.5}. Systems evolving near EP exhibit plentiful physical phenomena including non-reciprocal topological energy transfer\cite{RN1.2.1,RN1.6.1} and EP based devices\cite{RN1.7,RN3.1.6,RN3.1.7} while evolving DP only results in a geometry phase\cite{RN1.3.1}. In particular dynamically encircling around EP\cite{RN1.8.1} promise the opportunities to realize asymmetric mode switching\cite{RN1.2.1,RN1.9.2}, which has been exhibited in various system such as waveguides\cite{RN1.9.1}, circuits\cite{RN1.10}, and plasmonics\cite{RN1.11,RN1.11.2}. Due to the ability of enhancing light and matter interaction in an ultra-small volume, high quality microcavities\cite{RN3.1.8} have been promising platform to manipulate EP to approach state conversion\cite{RN1.13.1,RN1.13.2}, which has been experimentally demonstrated\cite{RN1.6.1}. Determined by orientation, location, and initial state, dynamically encircling EP presents outstanding robustness\cite{RN1.2.1} and chirality\cite{RN1.14}.

For the experimental realization of EP in the microcavity optical mode, the direct coupling of the two optical cavities is usually used\cite{RN2.1.1}.
According to previous research, only the laser detuning and the optical-optical coupling strength can be employed to tune in the whole system\cite{RN2.2.1,RN2.2.2} while there is difficulty in experiment to control the coupling strength between optical modes. The lack of control measurements sets an obstacle to dynamically encircling the EP and the extra degrees of parameter control are needed. To solve the problem, optomechanical system\cite{RN2.3.1,RN2.6.1,RN3.1.11,RN3.1.12,RN3.1.13,RN3.1.14,RN3.1.15,RN2.6.2,RN2.6.3,RN3.1.4,RN3.1.16}, where an auxiliary mechanical mode can be introduced to achieve indirect coupling between the two optical modes, has been proposed and experimentally realized recently\cite{RN3.1.9,RN2.5.1,RN3.1.10}. Through indirect coupling, the methods of regulating EP become more abundant.

In this paper, in order to enrich the manipulations of optical systems in dynamically encircling EP, we propose an largely detuned multimode optomechanical system to reach strong coupling regime with the aid of optomechanical dark mode. The effect of the initial state, orientation, location and the velocity in the time-dependent system dynamic evolution is investigated, whose track is a closed loop in the parameters space. The results show that state conversion may happen whether the path encloses the EP or not because of the effect of non-Hermiticity-induced nonadiabatic transition (NAT)\cite{RN3.1.1,RN3.1.2} while chirality conversion will appear if the path encloses the EP. Moreover, the nonreciprocity and the chirality of the topological energy transform efficiency depends on the locations of the control loops in the parameters space. Our work may inspire further experiments about photon states manipulation in large detuning domain of multimode optomechanical system and the devices based on extended non-Hermitian photonic architectures.

This article is organized as follows: In Sec.\ref{MODEL AND HAMILTONIAN}, we demonstrate the basic model and hamiltonian of the system. We study the dynamic evolution of the system in Sec.\ref{DYNAMICALLY}. We research the topological energy transfer in Sec.\ref{Topological}. Conclusion is given in Sec.\ref{Conclusion}. Appendix.\ref{AppendixA} presents how to choose independent variables of the control loop.

\section{MODEL AND HAMILTONIAN \label{MODEL AND HAMILTONIAN}}

\begin{figure}
	\centering
	\includegraphics[width=\linewidth]{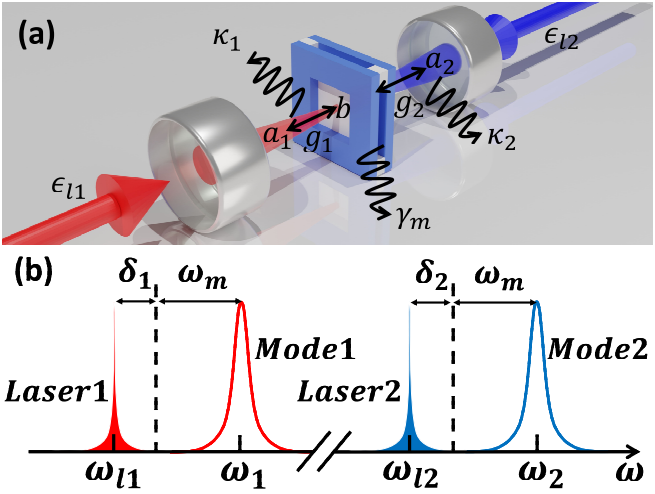}
	\caption{(a) Schematic of the optomechanical system composed of two optical modes ($a_1$ and $a_2$) and one mechanical mode ($b$). The two optical modes are coupled with the mechanical modes simultaneously. (b) Frequency spectrum of the two optical systems and two strong power lasers with the red detunings.}
	\label{model}
\end{figure}

The schematic of our system is shown in Fig.\ref{model}(a), composed of two optical modes and one mechanical mode. The mechanical mode can be a film, which is fixed left and right. The two optical modes are driven by strong lasers with central frequencies of $\omega_{l1}$ and $\omega_{l2}$. The damping (gaining) rates of the optical modes are $\kappa_j$ and the mechanical mode is $\gamma$. The mechanical mode is simultaneously coupled with two optical modes dispersively. The system hamiltonian can be described by

\begin{align}
    H &= H_{free} + H_{int} + H_{drive},
    \label{equation 1}
\end{align}
where 

\begin{align}
    H_{free} =& \omega_1{a_1}^\dagger a_1+\omega_2{a_2}^\dagger a_2+\omega_mb^\dagger b, \nonumber \\
    H_{int} =& g_1{a_1}^\dagger a_1(b^\dagger+b)+g_2{a_2}^\dagger a_2(b^\dagger+b), \nonumber\\
    H_{drive} =&i\sqrt{\kappa_{ex1}}\epsilon_{l1}e^{-i\omega_{l1}t}{a_1}^\dagger+i\sqrt{\kappa_{ex2}}\epsilon_{l2}e^{-i\omega_{l2}t}{a_2}^\dagger+H.c.,\nonumber\\
   	\label{equation 2}
\end{align}
$\omega_{j}$ denotes resonance frequency of the $j^{th}$ optical mode and $\omega_m$ describes the resonance frequency of the mechanical mode. $a_j$ and $b$ are the annihilation of the $j^{th}$ optical mode and mechanical mode respectively. $g_j$ describes the single photon optomechanical coupling rate and $\kappa_{exj}$ describes the damping rate of the couple between fibre and optical mode. $\epsilon_{lj}$ is the power of the $j^{th}$ laser.

In the interaction picture of the driving filed, the hamiltonian is changed to

\begin{align}
	H &= \omega_mb^\dagger b-\sum_{j=1,2}[\Delta_j{a_j}^\dagger a_j-g_j{a_j}^\dagger a_j(b^\dagger+b)] \nonumber \\ 
	  &+i\sqrt{\kappa_{ex1}}\epsilon_{l1}{a_1}^\dagger+i\sqrt{\kappa_{ex2}}\epsilon_{l2}{a_2}^\dagger,
	\label{equation 3}
\end{align}
$\Delta_j=\omega_{lj}-\omega_j$ is the detuning between the optical modes and the corresponding driving fields. It is convenient to solve the nonlinear Heisenberg equations if the hamiltonian above has been linearized standardly. The annihilations of the optical modes can be changed via $a_j\rightarrow\overline{a_j}+\delta a_j$. The linearized hamiltonian is turn to

\begin{align}
	H=&\omega_mb^\dagger b-\sum_{j=1,2}[\Delta_j{a_j}^\dagger a_j-G_j({a_j}^\dagger+a_j)(b^\dagger+b)],
	\label{equation 4}
\end{align}
$G_j=g_j\overline{a_j}$ describes the optically driven coupling between the mechanical mode and cavity mode j. $\overline{a_j}=\sqrt{\kappa_{exj}}\epsilon_{lj}/(-i\Delta_j+\frac{\kappa_j}{2})$ is the average of the annihilation. In the following, the strong-coupling regime, i.e.,$G_j>(\kappa_j,\gamma_m)$ and the typical limit $\kappa_j\gg\gamma_m$ is focused.

The case both the optical modes are driven under the red sidebands is considered. For convenience, the detuning can be turned to $\delta_j=-\Delta_j-\omega_m$. Under the condition of $\omega_m\gg(\delta_j,G_j)$ and the rotating-wave approximation, the hamiltonian can be written as $H_A=\sum_{j=1,2}[\delta_j{a_j}^\dagger a_j+G_j({a_j}^\dagger b+a_jb^\dagger)]$ in the interaction picture of $\omega_m(b^\dagger b+\sum_{j=1,2}{a_j}^\dagger a_j)$. For large detuning conditions $\delta_j\gg G_j$, the mechanical modes can be eliminated\cite{RN2.5.1}. We get the effective hamiltonian

\begin{align}
	\mathcal{H}=&
	\begin{pmatrix}
		\delta_1+\Omega_1-\frac{\kappa_1}{2}i & \Omega \\
		\Omega & \delta_2+\Omega_2-\frac{\kappa_2}{2}i
	\end{pmatrix},
	\label{equation 5}
\end{align}
$\Omega_j={G_j}^2/\delta_j$ is the resonances ac-Stark shift of cavity mode j. $\Omega=G_1G_2({\delta_1}^{-1}+{\delta_2}^{-1})/2$ describes the effective coupling between the two optical modes\cite{RN2.5.1}.

\begin{figure}
	\centering
	\includegraphics[width=\linewidth]{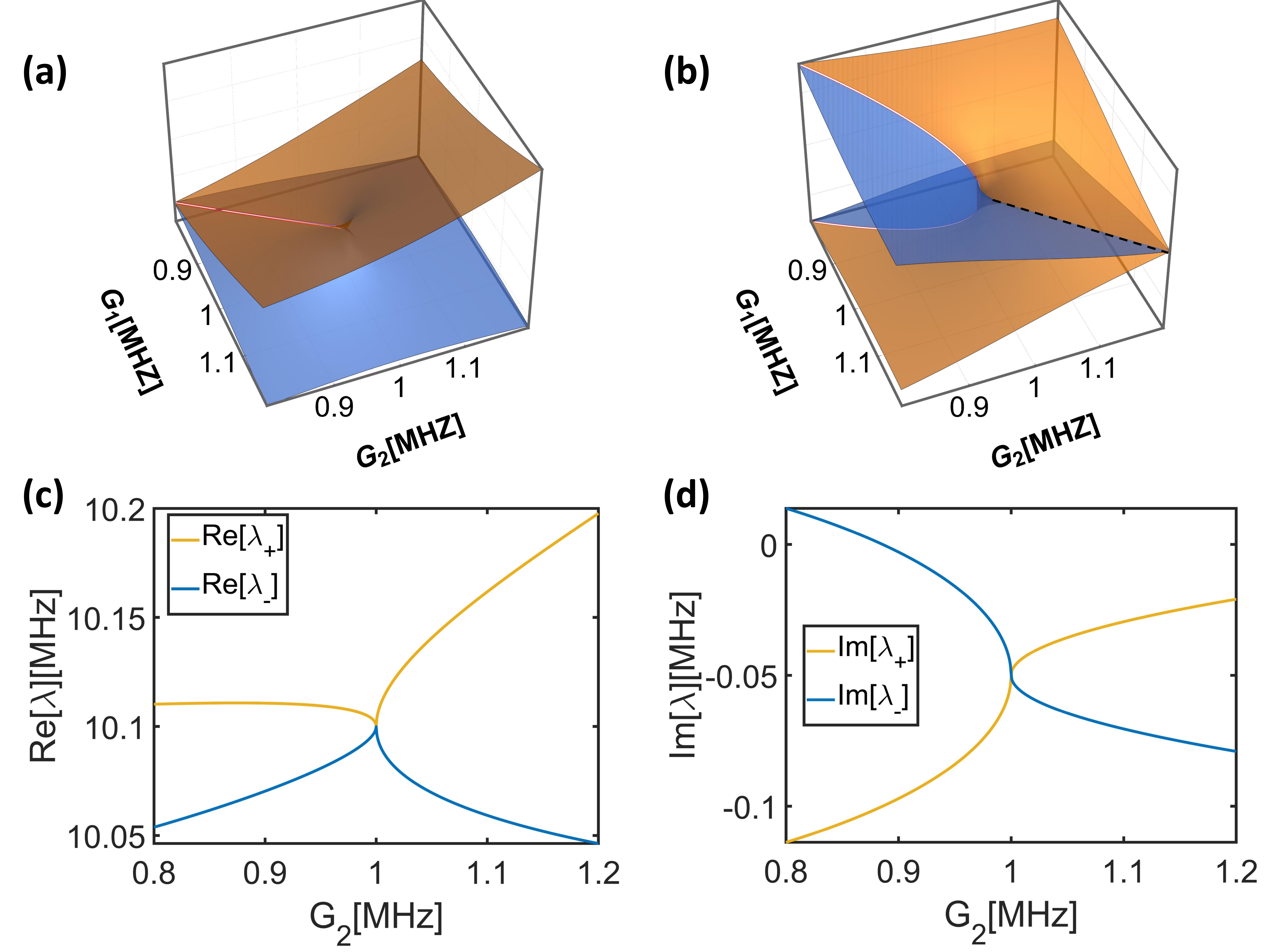}
	\caption{(a) Real part $Re[\lambda_\pm]$ and (b) Imaginary part $Im[\lambda_\pm]$ of the effective hamiltonian eigenvalues as functions of optomechanial coupling strengths $G_1$ and $G_2$. The black dashed line describes where the imaginary part of eigenvalues coalesce. (c) Real part $Re[\lambda_\pm]$ and (d) Imaginary part $Im[\lambda_\pm]$ of the effective hamiltonian eigenvalues as functions of optomechanial coupling strength $G_2$ with $G_1$ fixing at G. Colouring is chosen that yellow (blue) corresponds to the eigenvalue $\lambda_+$($\lambda_-$). $\delta_1=\delta_2=10G$, $\kappa_1=0.3G$, $\kappa_2=-0.1G$ are the parameters of the effective hamiltonian.}
	\label{hamiltonian}
\end{figure}

The eigenvalues of the effective hamiltonian in Eq.\ref{equation 5} can be obtained

\begin{align}
	\lambda_{\pm}=\frac{1}{2}(A_1+A_2)\pm\frac{\sqrt{(A_1-A_2)^2+4\Omega^2}}{2},
	\label{equation 9}
\end{align}
where $A_j=\delta_j+\Omega_j-\frac{\kappa_j}{2}i$. It is obvious to find that the effective hamiltonian in Eq.\ref{equation 5} has an EP if the conditions $\delta_1+\Omega_1=\delta_2+\Omega_2$ and $\Omega=\left|\kappa_1-\kappa_2\right|/4$ are satisfied. If the conditions that loss and gain of the two optical modes are balanced $\kappa_1=-\kappa_2$, and $\delta_1+\Omega_1=\delta_2+\Omega_2$ are satisfied, the effective hamiltonian in Eq.\ref{equation 5} will exhibit $\mathcal{PT}$ symmetry.

The real and imaginary part of the eigenvalues are shown in Fig.\ref{hamiltonian}(a) and Fig.\ref{hamiltonian}(b). $G_1$ and $G_2$ are chosen to be the independent variables, which can be tuned by changing the powers of the two driving lasers. How to choose these two independent variables will be presented at length in Appendix.\ref{AppendixA}. In short, this set of in dependent parameters is the best choice to show the structure of the effective hamiltonian. It is easy to be calculated that there is an EP when $G_1=G_2=G$. The two eigenvalues $\lambda_+$ and $\lambda_-$ coalesce at the EP. In the vicinity around the EP, the real and imaginary part of eigenvalues exhibit the same structure as Riemann sheets of complex square-root function $z^{\frac{1}{2}}$. On the basis of Fig.\ref{hamiltonian}(a) and Fig.\ref{hamiltonian}(b), $G_1$ is fixed at $G$ and the only independent variable is $G_2$. Real part and imaginary part of effective hamiltonian eigenvalues as function of $G_2$ are shown in Fig.\ref{hamiltonian}(c) and Fig.\ref{hamiltonian}(d) respectively. In these two subgraphs, it is convenient to see that when $G_2\neq G$ real parts and imaginary parts of eigenvalues are distinct while $G_2=G$ real parts and imaginary parts of eigenvalues are identical. 

\section{DYNAMICALLY ENCIRCLING AN EXCEPTIONAL POINT \label{DYNAMICALLY}}

The dynamic evolution of the operators depend on the Heisenberg equation $\frac{dA}{dt}=\frac{i}{h}[H,A]$. Because of the special structure of the Riemann sheets of the Fig.\ref{hamiltonian}.(a) and Fig.\ref{hamiltonian}.(b), when the $G_1$ and $G_2$ varied around an closed loop, there will be different evolution results depending on whether the loop encloses the EP. Curiously, the evolution results do not only depend on whether the loop encloses the EP, but also on the orientation of the loop, the initial state of the loop and the velocity of the evolution.

To realize the dynamic evolution of the hamiltonian in the parametric space, $G_1$ and $G_2$ will be tuned by changing the powers of the two input lasers $\epsilon_{l1}$ and $\epsilon_{l2}$. If we denotes $|\lambda_+\rangle$ and $|\lambda_-\rangle$ the instantaneous eigenvectors of the time-dependent hamiltonian, the time-dependent state can be expressed to $|\psi(t)\rangle=c_+(t)|\lambda_+(t)\rangle+c_-(t)|\lambda_-(t)\rangle$, where the $c_+$ and $c_-$ denote the amplitudes of the instantaneous eigenvalues. Because of the non-Hermitian hamiltonian, the instantaneous eigenvectors are not orthogonal, that is $\langle\lambda_-|\lambda_+\rangle\neq0$. So the amplitudes of instantaneous eigenvalues can not be obtained directly. That can be expressed by $c_+(t)\neq\langle\lambda_+(t)|\psi(t)\rangle$ and $c_-(t)\neq\langle\lambda_-(t)|\psi(t)\rangle$. To obtain the amplitudes, new vectors $|l_\pm(t)\rangle=|\lambda{\pm}(t)\rangle-\langle\lambda{\mp}(t)|\lambda{\pm}(t)\rangle|\lambda{\mp}(t)\rangle$ can be established, then the amplitudes can be determined by projecting the state after evolution onto the new vectors $c_+(t)=\frac{\langle l_+(t)|\psi(t)\rangle}{1-|\langle\lambda_-(t)|\lambda_+(t)\rangle|^2}$, $c_-(t)=\frac{\langle l_-(t)|\psi(t)\rangle}{1-|\langle\lambda_-(t)|\lambda_+(t)\rangle|^2}$\cite{RN3.1.3}.

\begin{figure*}
	\centering
	\includegraphics[width=\linewidth]{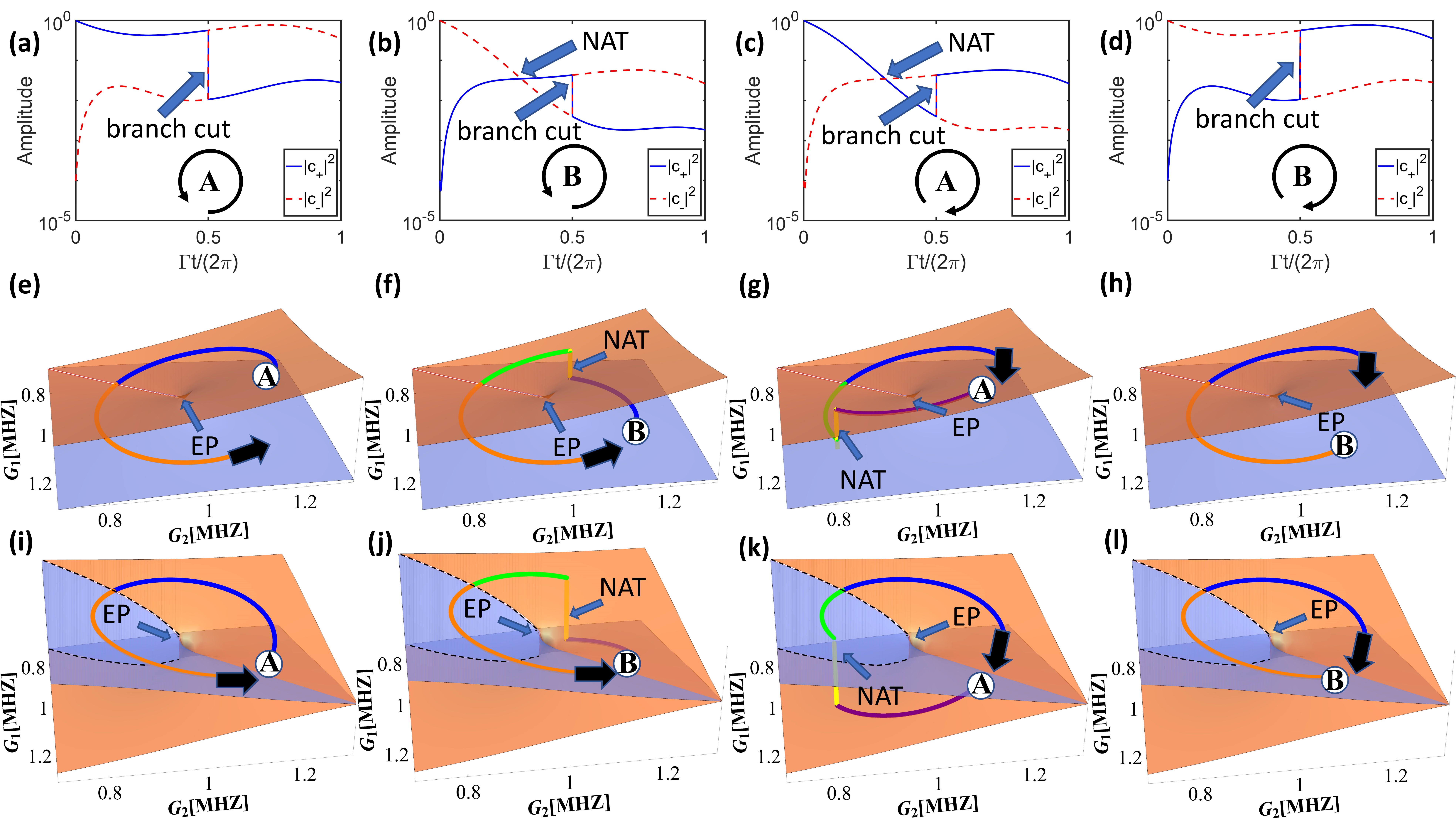}
	\caption{(a)-(d) The instantaneous amplitudes of eigenstates during the loop evolution, which encloses the EP. The blue solid line describes the amplitude of the $\lambda_+$ while the red dashed line describes the amplitude of the $\lambda_-$. (a) and (b) are the case the loop orientation is CCW while (c) and (d) are the case the loop orientation is CW. Point A(B) corresponds the case that the initial state is nearly $|\lambda_+\rangle$($|\lambda_-\rangle$). (a) and (c) are the case the initial state is A while (b) and (f) are the case the initial state is B. (e)-(h) Encircling path on the real part of the eigenstates of the hamiltonian corresponding to (a)-(d). (i)-(l) Encircling path on the imaginary part of the eigenstates of the hamiltonian corresponding to (a)-(d). The Blue(Orange) Riemann sheet is imaginary part of $\lambda_-(\lambda_+)$. The black dashed lines denote the branch-cut of the two imaginary parts of the eigenvalues Riemann sheets. The colour of the path corresponds different stages of the evolution. The parameters of the effective hamiltonian is the same as these in Fig.\ref{hamiltonian}(b).}
	\label{Encircling1}
\end{figure*}

\begin{figure*}
	\centering
	\includegraphics[width=\linewidth]{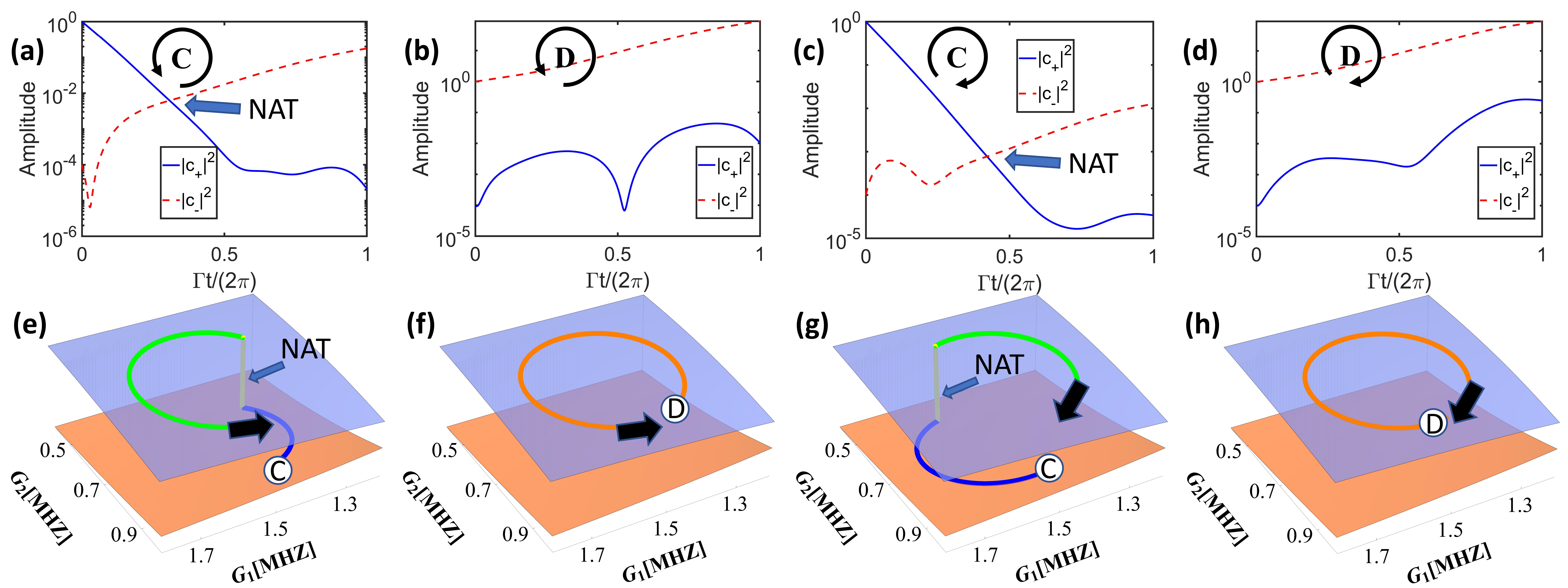}
	\caption{(a)-(d) The instantaneous amplitudes of eigenstates during the loop evolution, which do not enclose the EP. The blue solid line describes the amplitude of the $\lambda_+$ while the red dashed line describes the amplitude of the $\lambda_-$. (a) and (b) are the case the loop orientation is CCW while (c) and (d) are the case the loop orientation is CW. Point C(D) corresponds the case that the initial state is nearly $|\lambda_+\rangle$($|\lambda_-\rangle$). (a) and (c) are the case the initial state is A while (b) and (f) are the case the initial state is B. (e)-(h) Encircling path on the imaginary part of the eigenstates of the hamiltonian corresponding to (a)-(d). The Blue(Orange) Riemann sheet is imaginary part of $\lambda_-(\lambda_+)$. The colour of the path corresponds different stages of the evolution. The parameters of the effective hamiltonian is the same as these in Fig.\ref{hamiltonian}(b).}
	\label{Encircling2}
\end{figure*}

It is convenient to get the information of the evolution by observing the instantaneous amplitudes of the eigenvalues. The simplest function of the loop can be written by  $G_1(t)=\alpha_1 G_1^{EP}+\beta_1 G_1^{EP}cos(\Gamma t+\phi_0)$ and $G_2(t)=\alpha_2 G_2^{EP}+\beta_2 G_2^{EP}sin(\Gamma t+\phi_0)$. The system will obtain an EP when $G_i$ approaches $G_i^{EP}$. $\Gamma$ describes the velocity of evolution and the sign of $\Gamma$ describes the orientation of the loop. $\Gamma>0$ corresponds the loop is counterclockwise (CCW) while $\Gamma<0$ corresponds the loop is clockwise (CW). $\alpha_i$ determines the location of the loop center and $\beta_i$ determines the radius of the loop. The phase of starting point is determined by $\phi_0$ when we decide the time interval is $[0, \frac{2\pi}{|\Gamma|}]$. In the cases which loop encloses the EP, the starting point is chosen at the place where the imaginary of the eigenvalues coalesce. Without loss of generality, it is supposed that most of the energy is concentrated on one of the eigenstates initially, that is $c_+(0)\gg c_-(0)$ or $c_-(0)\gg c_+(0)$. These conditions above can correspond to initial time $t=0$ on the subgraphs in Fig.\ref{Encircling1}(a)-(d) and Fig.\ref{Encircling2}(a)-(d).

It is necessary to study the case that the loop encloses the EP and the evolution is adiabatic. $\alpha_1=\alpha_2=1$ and $\beta_1=\beta_2=0.2$ can make the loop enclose the EP. $\Gamma=0.1G$ ensure the evolution is adiabatic and $\phi_0=\frac{\pi}{4}$ make the starting point at the place where the imaginary of the eigenvalues coalesce. From Fig.\ref{Encircling1}(a) to Fig.\ref{Encircling1}(d), there is a state-flip when $\Gamma t/(\pi)=1$. It is a direct result because of the topological structure of the imaginary parts of eigenvalues around the EP. When the time is approaching $\Gamma t/(\pi)\rightarrow1$, if the state wants to keep unchanged, the imaginary part of the state will change abruptly due to the branch-cut, where connects two Riemann sheets of imaginary parts of eigenvalues. So the path will go to another Riemann sheet from the previous Riemann sheet via branch-cut, which is showed by the black dashed line in Fig.\ref{Encircling1}(i)-(l). While in Fig.\ref{Encircling1}(b) and Fig.\ref{Encircling1}(c), the NAT occurs, the state evolves to another sheet adiabatically. Because of the time-dependent hamiltonian, the state will not precisely be the eigenstate of the instantaneous hamiltonian during the evolution. If the state firstly propagates on the higher-loss sheet, the evolution will be unstable and the state will gradually propagate to the lower-loss sheet if the evolution is adiabatic. When $c_+(t_1)=c_-(t_1)$, we define it as the confirmation of appearance of the NAT at the time $t=t_1$\cite{RN3.1.2}. The yellow part of the path in Fig.\ref{Encircling1}(j)-(k) exhibit the NAT during the evolution. In Fig.\ref{Encircling1}(b), the state firstly propagate on the blue sheet which is higher-loss now, after a time delay the state will propagate to the orange sheet which is lower-loss now. While in Fig.\ref{Encircling1}(c), although the initial state is different from the case in Fig.\ref{Encircling1}(b), the orientation of the loop is opposite against that in Fig.\ref{Encircling1}(b) and there is also an NAT.

It is easy to find that if the loop orientation is CCW, the final state will evolve to $|\lambda_-\rangle$ whatever the initial state is. If the loop orientation is CW, the final state will evolve to $|\lambda_+\rangle$ whether the initial state is $|\lambda_+\rangle$ or $|\lambda_+\rangle$. We can also find that whether the final state is different from the initial state corresponds to whether the NAT has happened. 

To summarize, the results after evolution of the loop encloses the EP will be determined by the orientation of the loop and the initial state. Due to the different evolution orientation corresponds to different evolution result, there is chirality in the state conversion.

The case that loop does not enclose the EP is needed to compare with the case above. The evolution is also adiabatic. $\alpha_1=1.5,\alpha_2=0.5$ and $\beta_1=\beta_2=0.2$ make the loop does not enclose the EP. $\Gamma=0.1G$ ensure the evolution is adiabatic. Fig.\ref{Encircling2}(a) and Fig.\ref{Encircling2}(c) show that there is a state-flip during the loop evolution while there is not state-flip in Fig.\ref{Encircling2}(b) and Fig.\ref{Encircling2}(d). In Fig.\ref{Encircling2}(a) and Fig.\ref{Encircling2}(c) the initial state evolve on a higher-loss sheet firstly, NAT will happen after a time delay so the final state will be different from the initial state. In Fig.\ref{Encircling2}(b) and Fig.\ref{Encircling2}(d), the state keeps in the lower-loss sheet and there is not NAT happening. No matter what the initial state is, the final state must be the state on the lower-loss Riemann sheet, which is $|\lambda_-\rangle$ on this case. 

To summarize, if the loop does not enclose the EP, there will be not state-flip because of the topological structure of the Riemann sheet around the EP and there is not chirality in the state conversion. But due to the NAT effect, there may be state conversion if the initial state was initially on the higher-loss Riemann sheet.

\section{Topological Energy Transfer \label{Topological}}

From Fig.\ref{Encircling1} and Fig.\ref{Encircling2}, it is evident that the energy of the system is lost or gained during the loop evolution. This phenomena show that the total damping rate of the system may be positive or be negative depending on the sign of the imaginary parts of the instantaneous eigenvalues. We have study about the dynamic evolution under different conditions about the loop orientation, initial state and whether the loop encloses the EP. Here we want to research about the inflection of the loop velocity on the evolution. The evolutions in Fig.\ref{Encircling1} and Fig.\ref{Encircling2} are adiabatic, whose the period of the loop $\tau$ should satisfy $\tau\gg1/|\lambda_+-\lambda_-|$\cite{RN1.6.1}. Whether the loop is adiabatic determines the result of energy transfer. 

To concentrate our mind on the energy transfer during the loop, we focus on the relative energy before and after the evolution. When most of the energy is contributed on the state $|\lambda_+(0)\rangle$, one can define energy transfer efficiency $E_+=\frac{|c_-(\tau)|^2}{|c_+(\tau)|^2+|c_-(\tau)|^2}$. While most of the energy is contributed on the state $|\lambda_-(0)\rangle$, the energy transfer efficiency is $E_-=\frac{|c_+(\tau)|^2}{|c_+(\tau)|^2+|c_-(\tau)|^2}$.

\begin{figure*}
	\centering
	\includegraphics[width=\linewidth]{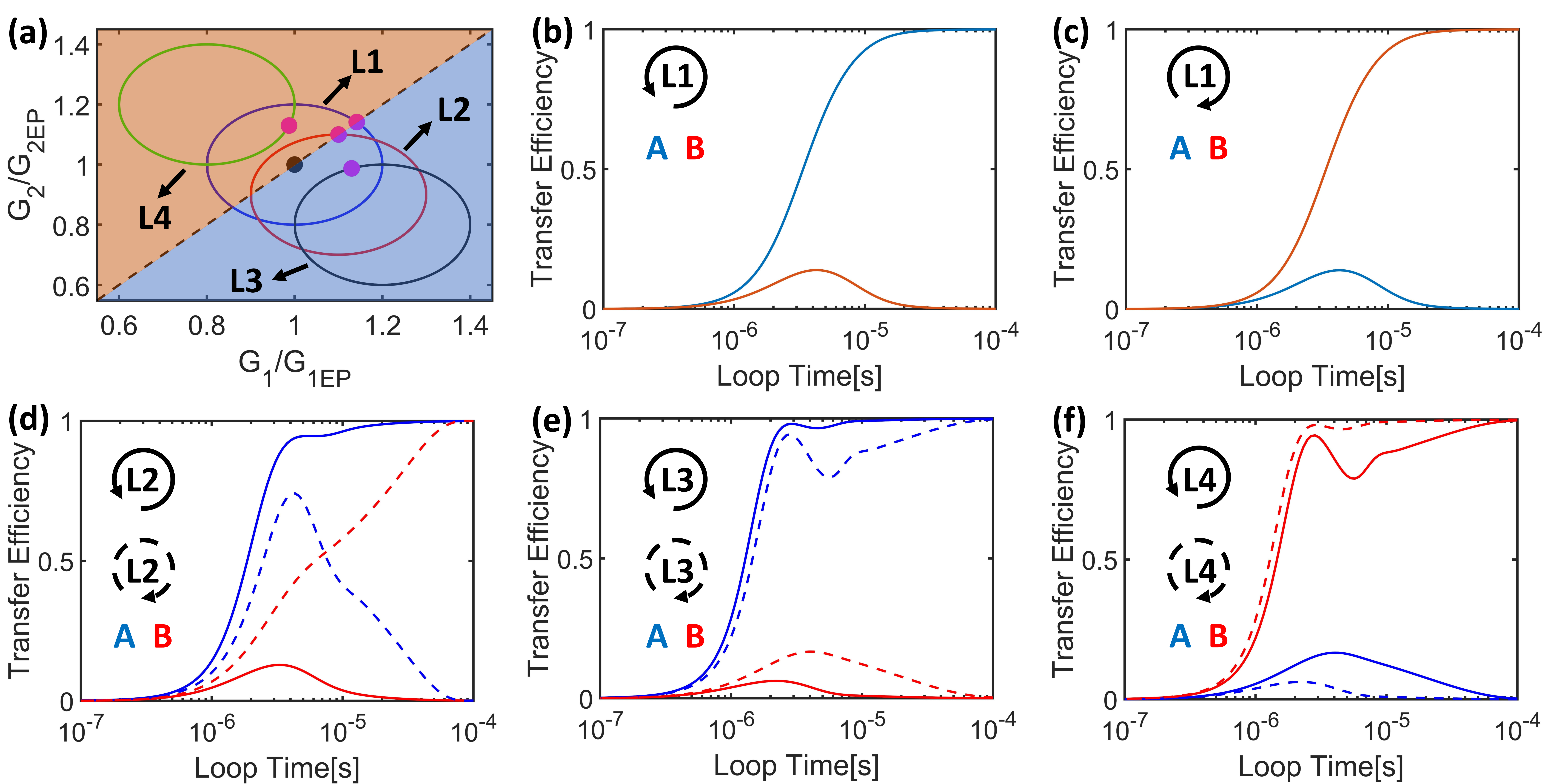}
	\caption{(a) Parametric space of the system. There are four different loops marked as L1-L4 in this subgraph. The black dashed line $G_1=G_2$ is the demarcation line for the less lossy region of the eigenvalues. The orange(blue) area represents the area where the damping rate of eigenvalue $\lambda_+(\lambda_-)$ is smaller. The black dot denotes the location of the EP and the magenta dots denote the starting points location. (b)-(c) Energy transfer efficiency E as a function of the period $\tau$ on the loop L1 described in (a). The blue(red) solid line is the transfer efficiency $E_+(E_-)$ where the state $|\lambda_+\rangle$($|\lambda_-\rangle$) initially be driven . That corresponds the initial state is at A(B). The loop orientation in (b) is CCW while in (c) is CW. (d)-(f) Energy transfer efficiency E as a function of the period $\tau$ on the loops L2-L4 described in (a) respectively. The blue(red) solid line is the transfer efficiency $E_+(E_-)$ where the initial state is at A(B) and the orientation is CCW. While the blue(red) dashed line is the transfer efficiency $E_+(E_-)$ which the initial state is at A(B) and the orientation is CW. The parameters of the effective hamiltonian is the same as these in Fig.\ref{hamiltonian}(b).}
	\label{TRANSFER1}
\end{figure*}

In Fig.\ref{TRANSFER1}(a), L1-L4 are four different loops to be investigated. The trajectory parameters of L1 is $\alpha_1=1,\alpha_2=1$, L2 is $\alpha_1=1.1,\alpha_2=0.9$, L3 is $\alpha_1=1.2,\alpha_2=0.8$ and L4 is $\alpha_1=0.8,\alpha_2=1.2$. $\beta_1=\beta_2=0.2$ is in L1-L4. The $\phi_0$ in L1-L2 ensures that the starting points at the place where the imaginary of the eigenvalues coalesce. In Fig.\ref{TRANSFER1}(b)-(f), when the period of the loop is small enough ($\tau\rightarrow0$), no matter which state is initially driven and what the orientation is, the transfer efficiency is negligible ($E\rightarrow0$). This can be explained by the system experienced a sudden perturbation which is not enough to switch the state to another. 

Fig.\ref{TRANSFER1}(b)-(d) correspond the cases that loops enclose the EP. As the period of the loop increases, it is easy to find that the transfer efficiencies $E_+$ and $E_-$ both increase. When the loop orientation is CCW, the transfer efficiency $E_+$ will increase to 1 as the loop evolution is adiabatic ($\tau\gg1\mu s$) while the transfer efficiency $E_-$ will go to zero after a maximum. When the loop orientation is CW, the transfer efficiency $E_-$ will increase to 1 as the loop evolution is adiabatic ($\tau\gg1\mu s$) while the transfer efficiency $E_+$ will go to zero after a maximum. Rapid evolution causes vanishing transfer efficiency while the adiabatic evolution results in a different transfer efficiency which depend on the loop orientation. These phenomena reflect the non-reciprocity of each topological operation.

Fig.\ref{TRANSFER1}(e)-(f) correspond the cases that loops do not enclose the EP. As the period of the loop increases, it is easy to find that the transfer efficiencies $E_+$ and $E_-$ both increase. When the loop is L3, the transfer efficiency $E_+$ will increase to 1 as the loop evolution is adiabatic ($\tau\gg1\mu s$) while the transfer efficiency $E_-$ will go to zero after a maximum. When the loop is L4, the transfer efficiency $E_-$ will increase to 1 as the loop evolution is adiabatic ($\tau\gg1\mu s$) while the transfer efficiency $E_+$ will go to zero after a maximum. It is obviously that whether the transfer efficiency increases to 1 depends on the location of the loop center, no matter what the loop orientation is. This behaviour in the cases that loops do not enclose the EP is different that the case that loops enclose the EP.

In Fig.\ref{TRANSFER1}(b)-(c), the loop L1 is symmetrical about the black dashed line, which is the demarcation line for the less lossy region of the eigenvalues. It is easy to find the transfer efficiency behaviour show symmetry. The blue(red) line in Fig.\ref{TRANSFER1}(b) is the same as the red(blue) line in Fig.\ref{TRANSFER1}(c). In Fig.\ref{TRANSFER1}(d), the loop do not show symmetrical about the black dashed line. The transfer efficiency in CW and CCW do not exhibit symmetry. It is obvious that the transfer efficiency $E_+$ in L2 increases quicker than that in the loop L1 when the orientation is CCW. When the orientation is CW, the maximum of the $E_+$ in L2 is larger than that in the loop which is symmetrical. These behaviours can be explained by that the time when the loop L2 goes through the region, where the gain of $|\lambda_-\rangle$ is larger, is longer than the time during on the region where $|\lambda_+\rangle$ is less lossy. It will cause the asymmetric in the transfer efficiency. In Fig.\ref{TRANSFER1}(e)-(f), the adiabatic evolution behaviour is mainly determined by the position of the center of the trajectory, and has little to do with the orientation of the trajectory. That can be explained that because the loop do not contain the EP, the topological structure does not cause state conversion but only the NAT causes the state conversion. The result of the evolution is not determined by the orientation but the location and the velocity of the loop. L3(L4) is completely in the region that $\lambda_-(\lambda_+)$ is less lossy so that $E_+(E_-)$ will approach 1 when the evolution is adiabatic. 

To summarize, if the loop encloses the EP, the behaviour of the energy transfer efficiency not only depend on the orientation of the loop but also the location of the loop center. While the loop does not enclose the EP, the behaviour is mainly determined by the location of the loop.

\begin{figure*}
	\centering
	\includegraphics[width=\linewidth]{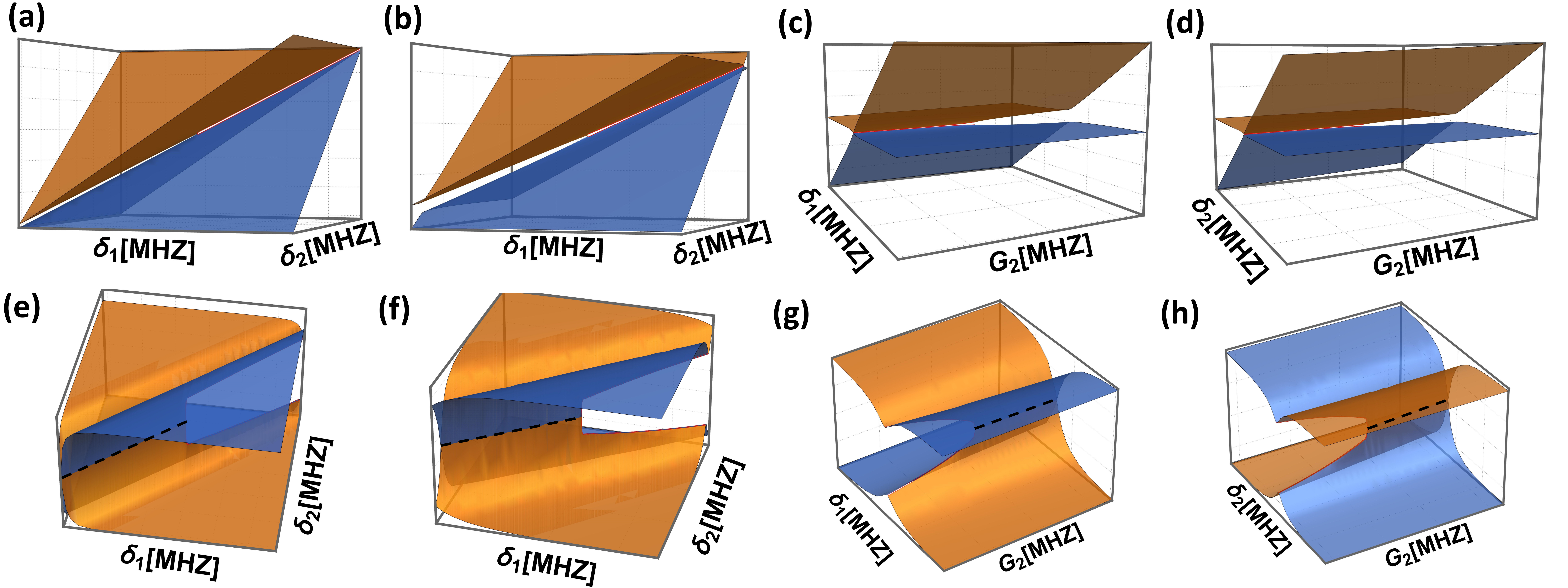}
	\caption{(a)-(d) Real part of the effective hamiltonian with different independent variable sets. (e)-(f) Imaginary part of the effective hamiltonian with different independent variable sets corresponding to (a)-(d). (a) $\delta_1$ and $\delta_2$ are chosen to be the independent variables.$G_1=G_2=G$,$\kappa_1=0.3G$,$\kappa_2=0.03G$. (b)$\delta_1$ and $\delta_2$ are chosen to be the independent variables.$G_1=G$,$G_2=5G$,$\kappa_1=0.3G$,$\kappa_2=1.5G$. (c)  $\delta_1$ and $\delta_2$ are chosen to be the independent variables.$G_1=G$,$\delta_2=4G$,$\kappa_1=0.6G$,$\kappa_2=0.2G$.(d)$\delta_1$ and $\delta_2$ are chosen to be the independent variables.$G_1=G$,$\delta_1=10G$,$\kappa_1=0.7G$,$\kappa_2=0.01G$. The black dashed line describes where the imaginary part of eigenvalues coalesce.}
	\label{APPENDIX1}
\end{figure*}

\section{Conclusion \label{Conclusion}}

We have demonstrated an optomechanical system, which is composed of two optical modes and one mechanical mode. This system will have an EP if some conditions are satisfied. Then we research the dynamic evolution of the time-dependent system, whose track is a closed loop in the parametric space. Whether there is chirality in the state conversion is determined by whether the loop encloses the EP, which reflects the topological structure around the EP influences the result of the dynamic evolution. We have also find that the initial state and the orientation affect the evolution result. At last, we study the impact of the evolution velocity and find that if the evolution is not adiabatic, the transfer efficiency is zero whatever the loop orientation is. While the evolution is adiabatic, the transfer efficiency depends on the loop orientation. We also find that the absolute location of the loop also have effect on the transfer efficiency. Our findings enrich the understanding of the closed path evolution of non-Hermitian system, which will be beneficial to the system control and broaden the mind of designing photonics devices based on EP. Combined with the results of this paper, there is great significance to investigate the dynamically encircling around the high-order EPs in the future. At the same time, using machine learning to learn the results of the control loop around the EPs under different trajectory parameters, it is hoped that the energy transfer efficiency can be improved at a larger velocity of the control loop.

\begin{acknowledgments}
	
This work is supported by the National Natural Science Foundation of China (61727801, and 62131002), National Key Research and Development Program of China (2017YFA0303700), Special Project for Research and Development in Key areas of Guangdong Province (2018B030325002), Beijing Advanced Innovation Center for Future Chip (ICFC), and Tsinghua University Initiative Scientific Research Program
	
\end{acknowledgments}

\appendix

\section{Selection of independent variables \label{AppendixA}}

In Sec.\ref{MODEL AND HAMILTONIAN}, we demonstrate the basic model and hamiltonian of the system and the dynamic evolution of the system is study in Sec.\ref{DYNAMICALLY}. Under the large detunings condition $\delta_j\gg G_j$, the mechanical modes can be eliminated so variables $G_j$, $\delta_j$ and $\kappa_j$ can be used as independent variables. When the couplings between fibre and optical modes are fixed, $\kappa_j$ is not convenient to regulated as an inherent property of the system material. If two independent variables are required, there are four variable combinations $(\delta_1,G_1)$ and $(\delta_2,G_2)$,$(\delta_1,G_2)$ and $(\delta_2,G_1)$ to choose. In Fig.\ref{hamiltonian} - \ref{Encircling2} $G_1$ and $G_2$ are chosen to be the independent variables considering the satisfaction of the detuning condition and the steepness of the imaginary parts of the eigenvalues.

Fig.\ref{APPENDIX1} shows real and imaginary parts of the effective hamiltonian with three different independent variable sets. In Fig.\ref{APPENDIX1}(e)-(f) $\delta_1$ and $\delta_2$ are chosen. It is easy to find that imaginary part around the black dashed line in Fig.\ref{APPENDIX1}(e) is steeper than that in Fig.\ref{APPENDIX1}(f). But under other non-independent variables in Fig.\ref{APPENDIX1}(f), the large detuning conditions that lead to the disappearance of the mechanical mode are more difficult to realize than that in Fig.\ref{APPENDIX1}(e). By contrast, imaginary part around the black dashed in Fig.\ref{APPENDIX1}(e)-(f) is steeper than that in Fig.\ref{APPENDIX1}(g)-(h). In the last four subplots, the independent variables set of Fig.\ref{APPENDIX1}(g) is chosen as the optimal solution. If Fig.\ref{hamiltonian}(b), whose independent variables set is $(G_1,G_2)$, is also taken into account, we can find that imaginary part around the black dashed line in Fig.\ref{hamiltonian}(b) is the flattest and the conditions are easier to realize. On the other hand, where the black dashed line corresponds to the imaginary part in the parameters space, the real parts of the eigenvalues are inconsistent. In Fig.\ref{APPENDIX1}(a)-(d), it is shown that the more flat around the black dashed line in Fig.\ref{APPENDIX1}(e)-(h), the difference of real part of eigenvalues is larger in the same place of the parameter place.

\begin{figure}
	\centering
	\includegraphics[width=\linewidth]{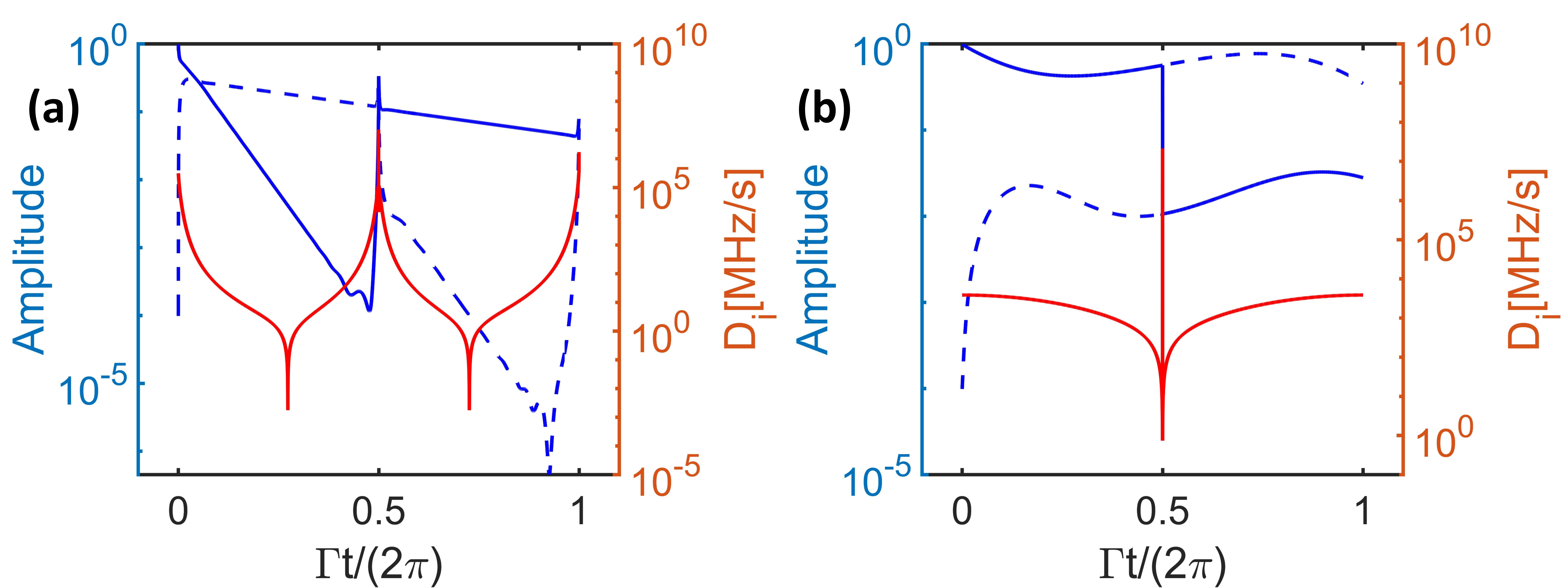}
	\caption{The instantaneous amplitudes of eigenstates and differential of imaginary part of effective hamiltonian $D_i$ during the loop evolution with two different independent variable sets. The evolution loop and other non-independent parameters are the same as in Fig.\ref{TRANSFER1}(a). Blue solid(dashed) line describes the amplitude of the $\lambda_+$($\lambda_-$). Red solid line describes the differential of the imaginary part of $\lambda_+$($\lambda_-$). (a) Variable set is $(\delta_1,\delta_2)$ and other non-independent parameters are the same as in Fig.\ref{APPENDIX1}(a). (b)Variable set is $(G_1,G_2)$ and other non-independent parameters are the same as in Fig.\ref{hamiltonian}(b).}
	\label{APPENDIX2}
\end{figure}

In order to study the influence about the steepness of the imaginary part of the effective hamiltonian to the evolution, we define the differential of imaginary part of eigenvalues  $D_i=\frac{Im[\lambda](i+1)-Im[\lambda](i)}{dt}$. $dt$ denotes the time interval of $Im[\lambda](i+1)$ and $Im[\lambda](i)$, which is a constant here. Fig.\ref{APPENDIX2} shows the instantaneous amplitudes of eigenstates and differential of Im$[\lambda]$ during the loop evolution with variable sets $(\delta_1,\delta_2)$ and $(G_1,G_2)$. It is easy to find that if $D_i$ is large enough, it can cause state conversion. In Fig.\ref{APPENDIX2}(a), we can find that there is a state conversion when $\Gamma t/(\pi)\rightarrow1$, which corresponds to the branch cut of the topological structure of the Riemann sheet and the magnitude of $D_i$ is $10^7$. The amplitudes of $\lambda_\pm$ tends to approach each other when $\Gamma t/(\pi)\rightarrow2$, correspond to the magnitude of $D_i$ here is $10^6$, which is large enough here. This phenomena corresponds to the imaginary part around the black dashed line in Fig.\ref{APPENDIX1}(a) is steepest in the last four subgraphs in Fig.\ref{APPENDIX1}. When $\Gamma t/(\pi)\rightarrow2$ in Fig.\ref{APPENDIX2}(b), it is obviously to find that the magnitude of $D_i$ here is $10^3$, which is smaller enough that there is not a state conversion. Comprehensive steepness and whether the large detunings conditions are easy to satisfy, variable sets $(G_1,G_2)$ is chosen to be the independent variables.

%

%\bibliographystyle{apsrev4-2.bst}

%\bibliography{sample.bib}

\begin{thebibliography}{46}%
	\makeatletter
	\providecommand \@ifxundefined [1]{%
		\@ifx{#1\undefined}
	}%
	\providecommand \@ifnum [1]{%
		\ifnum #1\expandafter \@firstoftwo
		\else \expandafter \@secondoftwo
		\fi
	}%
	\providecommand \@ifx [1]{%
		\ifx #1\expandafter \@firstoftwo
		\else \expandafter \@secondoftwo
		\fi
	}%
	\providecommand \natexlab [1]{#1}%
	\providecommand \enquote  [1]{``#1''}%
	\providecommand \bibnamefont  [1]{#1}%
	\providecommand \bibfnamefont [1]{#1}%
	\providecommand \citenamefont [1]{#1}%
	\providecommand \href@noop [0]{\@secondoftwo}%
	\providecommand \href [0]{\begingroup \@sanitize@url \@href}%
	\providecommand \@href[1]{\@@startlink{#1}\@@href}%
	\providecommand \@@href[1]{\endgroup#1\@@endlink}%
	\providecommand \@sanitize@url [0]{\catcode `\\12\catcode `\$12\catcode
		`\&12\catcode `\#12\catcode `\^12\catcode `\_12\catcode `\%12\relax}%
	\providecommand \@@startlink[1]{}%
	\providecommand \@@endlink[0]{}%
	\providecommand \url  [0]{\begingroup\@sanitize@url \@url }%
	\providecommand \@url [1]{\endgroup\@href {#1}{\urlprefix }}%
	\providecommand \urlprefix  [0]{URL }%
	\providecommand \Eprint [0]{\href }%
	\providecommand \doibase [0]{http://dx.doi.org/}%
	\providecommand \selectlanguage [0]{\@gobble}%
	\providecommand \bibinfo  [0]{\@secondoftwo}%
	\providecommand \bibfield  [0]{\@secondoftwo}%
	\providecommand \translation [1]{[#1]}%
	\providecommand \BibitemOpen [0]{}%
	\providecommand \bibitemStop [0]{}%
	\providecommand \bibitemNoStop [0]{.\EOS\space}%
	\providecommand \EOS [0]{\spacefactor3000\relax}%
	\providecommand \BibitemShut  [1]{\csname bibitem#1\endcsname}%
	\let\auto@bib@innerbib\@empty
	%</preamble>
	\bibitem [{\citenamefont {Peng}\ \emph
		{et~al.}(2014{\natexlab{a}})\citenamefont {Peng}, \citenamefont {\"{O}zdemir},
		\citenamefont {Lei}, \citenamefont {Monifi}, \citenamefont {Gianfreda},
		\citenamefont {Long}, \citenamefont {Fan}, \citenamefont {Nori},
		\citenamefont {Bender},\ and\ \citenamefont {Yang}}]{RN2.1.1}%
	\BibitemOpen
	\bibfield  {author} {\bibinfo {author} {\bibfnamefont {B.}~\bibnamefont
			{Peng}}, \bibinfo {author} {\bibfnamefont {\c{S}~K.}\ \bibnamefont {\"{O}zdemir}},
		\bibinfo {author} {\bibfnamefont {F.}~\bibnamefont {Lei}}, \bibinfo {author}
		{\bibfnamefont {F.}~\bibnamefont {Monifi}}, \bibinfo {author} {\bibfnamefont
			{M.}~\bibnamefont {Gianfreda}}, \bibinfo {author} {\bibfnamefont {G.~L.}\
			\bibnamefont {Long}}, \bibinfo {author} {\bibfnamefont {S.}~\bibnamefont
			{Fan}}, \bibinfo {author} {\bibfnamefont {F.}~\bibnamefont {Nori}}, \bibinfo
		{author} {\bibfnamefont {C.~M.}\ \bibnamefont {Bender}}, \ and\ \bibinfo
		{author} {\bibfnamefont {L.}~\bibnamefont {Yang}},\ }\href {\doibase
		10.1038/nphys2927} {\bibfield  {journal} {\bibinfo  {journal} {Nature
				Physics}\ }\textbf {\bibinfo {volume} {10}},\ \bibinfo {pages} {394}
		(\bibinfo {year} {2014}{\natexlab{a}})}\BibitemShut {NoStop}%
	\bibitem [{\citenamefont {Chen}\ \emph {et~al.}(2017)\citenamefont {Chen},
		\citenamefont {Kaya~\"{O}zdemir}, \citenamefont {Zhao}, \citenamefont
		{Wiersig},\ and\ \citenamefont {Yang}}]{RN1.13.1}%
	\BibitemOpen
	\bibfield  {author} {\bibinfo {author} {\bibfnamefont {W.}~\bibnamefont
			{Chen}}, \bibinfo {author} {\bibfnamefont {\c{S}}~\bibnamefont {Kaya~\"{O}zdemir}},
		\bibinfo {author} {\bibfnamefont {G.}~\bibnamefont {Zhao}}, \bibinfo {author}
		{\bibfnamefont {J.}~\bibnamefont {Wiersig}}, \ and\ \bibinfo {author}
		{\bibfnamefont {L.}~\bibnamefont {Yang}},\ }\href {\doibase
		10.1038/nature23281} {\bibfield  {journal} {\bibinfo  {journal} {Nature}\
		}\textbf {\bibinfo {volume} {548}},\ \bibinfo {pages} {192} (\bibinfo {year}
		{2017})}\BibitemShut {NoStop}%
	\bibitem [{\citenamefont {Heiss}(2004)}]{RN1.1}%
	\BibitemOpen
	\bibfield  {author} {\bibinfo {author} {\bibfnamefont {W.~D.}\ \bibnamefont
			{Heiss}},\ }\href {\doibase 10.1088/0305-4470/37/6/034} {\bibfield  {journal}
		{\bibinfo  {journal} {Journal of Physics A: Mathematical and General}\
		}\textbf {\bibinfo {volume} {37}},\ \bibinfo {pages} {2455} (\bibinfo {year}
		{2004})}\BibitemShut {NoStop}%
	\bibitem [{\citenamefont {Doppler}\ \emph {et~al.}(2016)\citenamefont
		{Doppler}, \citenamefont {Mailybaev}, \citenamefont {B枚hm}, \citenamefont
		{Kuhl}, \citenamefont {Girschik}, \citenamefont {Libisch}, \citenamefont
		{Milburn}, \citenamefont {Rabl}, \citenamefont {Moiseyev},\ and\
		\citenamefont {Rotter}}]{RN1.2.1}%
	\BibitemOpen
	\bibfield  {author} {\bibinfo {author} {\bibfnamefont {J.}~\bibnamefont
			{Doppler}}, \bibinfo {author} {\bibfnamefont {A.~A.}\ \bibnamefont
			{Mailybaev}}, \bibinfo {author} {\bibfnamefont {J.}~\bibnamefont {B\"{o}hm}},
		\bibinfo {author} {\bibfnamefont {U.}~\bibnamefont {Kuhl}}, \bibinfo {author}
		{\bibfnamefont {A.}~\bibnamefont {Girschik}}, \bibinfo {author}
		{\bibfnamefont {F.}~\bibnamefont {Libisch}}, \bibinfo {author} {\bibfnamefont
			{T.~J.}\ \bibnamefont {Milburn}}, \bibinfo {author} {\bibfnamefont
			{P.}~\bibnamefont {Rabl}}, \bibinfo {author} {\bibfnamefont {N.}~\bibnamefont
			{Moiseyev}}, \ and\ \bibinfo {author} {\bibfnamefont {S.}~\bibnamefont
			{Rotter}},\ }\href {\doibase 10.1038/nature18605} {\bibfield  {journal}
		{\bibinfo  {journal} {Nature}\ }\textbf {\bibinfo {volume} {537}},\ \bibinfo
		{pages} {76} (\bibinfo {year} {2016})}\BibitemShut {NoStop}%
	\bibitem [{\citenamefont {Lai}\ \emph {et~al.}(2019)\citenamefont {Lai},
		\citenamefont {Lu}, \citenamefont {Suh}, \citenamefont {Yuan},\ and\
		\citenamefont {Vahala}}]{RN1.2.2}%
	\BibitemOpen
	\bibfield  {author} {\bibinfo {author} {\bibfnamefont {Y.-H.}\ \bibnamefont
			{Lai}}, \bibinfo {author} {\bibfnamefont {Y.-K.}\ \bibnamefont {Lu}},
		\bibinfo {author} {\bibfnamefont {M.-G.}\ \bibnamefont {Suh}}, \bibinfo
		{author} {\bibfnamefont {Z.}~\bibnamefont {Yuan}}, \ and\ \bibinfo {author}
		{\bibfnamefont {K.}~\bibnamefont {Vahala}},\ }\href {\doibase
		10.1038/s41586-019-1777-z} {\bibfield  {journal} {\bibinfo  {journal}
			{Nature}\ }\textbf {\bibinfo {volume} {576}},\ \bibinfo {pages} {65}
		(\bibinfo {year} {2019})}\BibitemShut {NoStop}%
	\bibitem [{\citenamefont {Berry}\ and\ \citenamefont
		{Wilkinson}(1984)}]{RN1.3.1}%
	\BibitemOpen
	\bibfield  {author} {\bibinfo {author} {\bibfnamefont {M.~V.}\ \bibnamefont
			{Berry}}\ and\ \bibinfo {author} {\bibfnamefont {M.}~\bibnamefont
			{Wilkinson}},\ }\href {\doibase 10.1098/rspa.1984.0022} {\bibfield  {journal}
		{\bibinfo  {journal} {Proceedings of the Royal Society of London. A.
				Mathematical and Physical Sciences}\ }\textbf {\bibinfo {volume} {392}},\
		\bibinfo {pages} {15} (\bibinfo {year} {1984})}\BibitemShut {NoStop}%
	\bibitem [{\citenamefont {Wiersig}(2014)}]{RN1.5.1}%
	\BibitemOpen
	\bibfield  {author} {\bibinfo {author} {\bibfnamefont {J.}~\bibnamefont
			{Wiersig}},\ }\href {\doibase 10.1103/PhysRevLett.112.203901} {\bibfield
		{journal} {\bibinfo  {journal} {Phys. Rev. Lett.}\ }\textbf {\bibinfo
			{volume} {112}},\ \bibinfo {pages} {203901} (\bibinfo {year}
		{2014})}\BibitemShut {NoStop}%
	\bibitem [{\citenamefont {Jing}\ \emph {et~al.}(2014)\citenamefont {Jing},
		\citenamefont {\"Ozdemir}, \citenamefont {L\"u}, \citenamefont {Zhang},
		\citenamefont {Yang},\ and\ \citenamefont {Nori}}]{RN1.4.3}%
	\BibitemOpen
	\bibfield  {author} {\bibinfo {author} {\bibfnamefont {H.}~\bibnamefont
			{Jing}}, \bibinfo {author} {\bibfnamefont {S.~K.}\ \bibnamefont {\"Ozdemir}},
		\bibinfo {author} {\bibfnamefont {X.-Y.}\ \bibnamefont {L\"u}}, \bibinfo
		{author} {\bibfnamefont {J.}~\bibnamefont {Zhang}}, \bibinfo {author}
		{\bibfnamefont {L.}~\bibnamefont {Yang}}, \ and\ \bibinfo {author}
		{\bibfnamefont {F.}~\bibnamefont {Nori}},\ }\href {\doibase
		10.1103/PhysRevLett.113.053604} {\bibfield  {journal} {\bibinfo  {journal}
			{Phys. Rev. Lett.}\ }\textbf {\bibinfo {volume} {113}},\ \bibinfo {pages}
		{053604} (\bibinfo {year} {2014})}\BibitemShut {NoStop}%
	\bibitem [{\citenamefont {Zhang}\ \emph {et~al.}(2018)\citenamefont {Zhang},
		\citenamefont {Peng}, \citenamefont {\"{O}zdemir}, \citenamefont {Pichler},
		\citenamefont {Krimer}, \citenamefont {Zhao}, \citenamefont {Nori},
		\citenamefont {Liu}, \citenamefont {Rotter},\ and\ \citenamefont
		{Yang}}]{RN1.4.1}%
	\BibitemOpen
	\bibfield  {author} {\bibinfo {author} {\bibfnamefont {J.}~\bibnamefont
			{Zhang}}, \bibinfo {author} {\bibfnamefont {B.}~\bibnamefont {Peng}},
		\bibinfo {author} {\bibfnamefont {\c{S}~K.}\ \bibnamefont {\"{O}zdemir}}, \bibinfo
		{author} {\bibfnamefont {K.}~\bibnamefont {Pichler}}, \bibinfo {author}
		{\bibfnamefont {D.~O.}\ \bibnamefont {Krimer}}, \bibinfo {author}
		{\bibfnamefont {G.}~\bibnamefont {Zhao}}, \bibinfo {author} {\bibfnamefont
			{F.}~\bibnamefont {Nori}}, \bibinfo {author} {\bibfnamefont {Y.-x.}\
			\bibnamefont {Liu}}, \bibinfo {author} {\bibfnamefont {S.}~\bibnamefont
			{Rotter}}, \ and\ \bibinfo {author} {\bibfnamefont {L.}~\bibnamefont
			{Yang}},\ }\href {\doibase 10.1038/s41566-018-0213-5} {\bibfield  {journal}
		{\bibinfo  {journal} {Nature Photonics}\ }\textbf {\bibinfo {volume} {12}},\
		\bibinfo {pages} {479} (\bibinfo {year} {2018})}\BibitemShut {NoStop}%
	\bibitem [{\citenamefont {Jiang}\ \emph {et~al.}(2018)\citenamefont {Jiang},
		\citenamefont {Maayani}, \citenamefont {Carmon}, \citenamefont {Nori},\ and\
		\citenamefont {Jing}}]{RN1.4.2}%
	\BibitemOpen
	\bibfield  {author} {\bibinfo {author} {\bibfnamefont {Y.}~\bibnamefont
			{Jiang}}, \bibinfo {author} {\bibfnamefont {S.}~\bibnamefont {Maayani}},
		\bibinfo {author} {\bibfnamefont {T.}~\bibnamefont {Carmon}}, \bibinfo
		{author} {\bibfnamefont {F.}~\bibnamefont {Nori}}, \ and\ \bibinfo {author}
		{\bibfnamefont {H.}~\bibnamefont {Jing}},\ }\href {\doibase
		10.1103/PhysRevApplied.10.064037} {\bibfield  {journal} {\bibinfo  {journal}
			{Phys. Rev. Applied}\ }\textbf {\bibinfo {volume} {10}},\ \bibinfo {pages}
		{064037} (\bibinfo {year} {2018})}\BibitemShut {NoStop}%
	\bibitem [{\citenamefont {Qin}\ \emph {et~al.}(2019)\citenamefont {Qin},
		\citenamefont {Wang}, \citenamefont {Wen}, \citenamefont {Ruan},\ and\
		\citenamefont {Long}}]{RN1.5.3}%
	\BibitemOpen
	\bibfield  {author} {\bibinfo {author} {\bibfnamefont {G.-Q.}\ \bibnamefont
			{Qin}}, \bibinfo {author} {\bibfnamefont {M.}~\bibnamefont {Wang}}, \bibinfo
		{author} {\bibfnamefont {J.-W.}\ \bibnamefont {Wen}}, \bibinfo {author}
		{\bibfnamefont {D.}~\bibnamefont {Ruan}}, \ and\ \bibinfo {author}
		{\bibfnamefont {G.-L.}\ \bibnamefont {Long}},\ }\href {\doibase
		10.1364/PRJ.7.001440} {\bibfield  {journal} {\bibinfo  {journal} {Photonics
				Research}\ }\textbf {\bibinfo {volume} {7}},\ \bibinfo {pages} {1440}
		(\bibinfo {year} {2019})}\BibitemShut {NoStop}%
	\bibitem [{\citenamefont {Wiersig}(2020)}]{RN1.5.2}%
	\BibitemOpen
	\bibfield  {author} {\bibinfo {author} {\bibfnamefont {J.}~\bibnamefont
			{Wiersig}},\ }\href {\doibase 10.1364/PRJ.396115} {\bibfield  {journal}
		{\bibinfo  {journal} {Photonics Research}\ }\textbf {\bibinfo {volume} {8}},\
		\bibinfo {pages} {1457} (\bibinfo {year} {2020})}\BibitemShut {NoStop}%
	\bibitem [{\citenamefont {Mao}\ \emph {et~al.}(2020)\citenamefont {Mao},
		\citenamefont {Qin}, \citenamefont {Yang}, \citenamefont {Zhang},
		\citenamefont {Wang},\ and\ \citenamefont {Long}}]{RN1.5.4}%
	\BibitemOpen
	\bibfield  {author} {\bibinfo {author} {\bibfnamefont {X.}~\bibnamefont
			{Mao}}, \bibinfo {author} {\bibfnamefont {G.-Q.}\ \bibnamefont {Qin}},
		\bibinfo {author} {\bibfnamefont {H.}~\bibnamefont {Yang}}, \bibinfo {author}
		{\bibfnamefont {H.}~\bibnamefont {Zhang}}, \bibinfo {author} {\bibfnamefont
			{M.}~\bibnamefont {Wang}}, \ and\ \bibinfo {author} {\bibfnamefont {G.-L.}\
			\bibnamefont {Long}},\ }\href {\doibase 10.1088/1367-2630/abaacd} {\bibfield
		{journal} {\bibinfo  {journal} {New Journal of Physics}\ }\textbf {\bibinfo
			{volume} {22}},\ \bibinfo {pages} {093009} (\bibinfo {year}
		{2020})}\BibitemShut {NoStop}%
	\bibitem [{\citenamefont {Qin}\ \emph {et~al.}(2021)\citenamefont {Qin},
		\citenamefont {Xie}, \citenamefont {Zhang}, \citenamefont {Hu}, \citenamefont
		{Wang}, \citenamefont {Li}, \citenamefont {Xu}, \citenamefont {Lei},
		\citenamefont {Ruan},\ and\ \citenamefont {Long}}]{RN1.5.5}%
	\BibitemOpen
	\bibfield  {author} {\bibinfo {author} {\bibfnamefont {G.-Q.}\ \bibnamefont
			{Qin}}, \bibinfo {author} {\bibfnamefont {R.-R.}\ \bibnamefont {Xie}},
		\bibinfo {author} {\bibfnamefont {H.}~\bibnamefont {Zhang}}, \bibinfo
		{author} {\bibfnamefont {Y.-Q.}\ \bibnamefont {Hu}}, \bibinfo {author}
		{\bibfnamefont {M.}~\bibnamefont {Wang}}, \bibinfo {author} {\bibfnamefont
			{G.-Q.}\ \bibnamefont {Li}}, \bibinfo {author} {\bibfnamefont
			{H.}~\bibnamefont {Xu}}, \bibinfo {author} {\bibfnamefont {F.}~\bibnamefont
			{Lei}}, \bibinfo {author} {\bibfnamefont {D.}~\bibnamefont {Ruan}}, \ and\
		\bibinfo {author} {\bibfnamefont {G.-L.}\ \bibnamefont {Long}},\ }\href
	{\doibase https://doi.org/10.1002/lpor.202000569} {\bibfield  {journal}
		{\bibinfo  {journal} {Laser \& Photonics Reviews}\ }\textbf {\bibinfo {volume}
			{15}},\ \bibinfo {pages} {2000569} (\bibinfo {year} {2021})}\BibitemShut {NoStop}%
	\bibitem [{\citenamefont {Yang}\ \emph {et~al.}(2021)\citenamefont {Yang},
		\citenamefont {Mao}, \citenamefont {Qin}, \citenamefont {Wang}, \citenamefont
		{Zhang}, \citenamefont {Ruan},\ and\ \citenamefont {Long}}]{RN3.1.5}%
	\BibitemOpen
	\bibfield  {author} {\bibinfo {author} {\bibfnamefont {H.}~\bibnamefont
			{Yang}}, \bibinfo {author} {\bibfnamefont {X.}~\bibnamefont {Mao}}, \bibinfo
		{author} {\bibfnamefont {G.-Q.}\ \bibnamefont {Qin}}, \bibinfo {author}
		{\bibfnamefont {M.}~\bibnamefont {Wang}}, \bibinfo {author} {\bibfnamefont
			{H.}~\bibnamefont {Zhang}}, \bibinfo {author} {\bibfnamefont
			{D.}~\bibnamefont {Ruan}}, \ and\ \bibinfo {author} {\bibfnamefont {G.-L.}\
			\bibnamefont {Long}},\ }\href {\doibase 10.1364/OL.435843} {\bibfield
		{journal} {\bibinfo  {journal} {Optics Letters}\ }\textbf {\bibinfo {volume}
			{46}},\ \bibinfo {pages} {4025} (\bibinfo {year} {2021})}\BibitemShut
	{NoStop}%
	\bibitem [{\citenamefont {Xu}\ \emph {et~al.}(2016)\citenamefont {Xu},
		\citenamefont {Mason}, \citenamefont {Jiang},\ and\ \citenamefont
		{Harris}}]{RN1.6.1}%
	\BibitemOpen
	\bibfield  {author} {\bibinfo {author} {\bibfnamefont {H.}~\bibnamefont
			{Xu}}, \bibinfo {author} {\bibfnamefont {D.}~\bibnamefont {Mason}}, \bibinfo
		{author} {\bibfnamefont {L.}~\bibnamefont {Jiang}}, \ and\ \bibinfo {author}
		{\bibfnamefont {J.~G.~E.}\ \bibnamefont {Harris}},\ }\href {\doibase
		10.1038/nature18604} {\bibfield  {journal} {\bibinfo  {journal} {Nature}\
		}\textbf {\bibinfo {volume} {537}},\ \bibinfo {pages} {80} (\bibinfo {year}
		{2016})}\BibitemShut {NoStop}%
	\bibitem [{\citenamefont {Miri}\ and\ \citenamefont {Al\`{u}}(2019)}]{RN1.7}%
	\BibitemOpen
	\bibfield  {author} {\bibinfo {author} {\bibfnamefont {M.-A.}\ \bibnamefont
			{Miri}}\ and\ \bibinfo {author} {\bibfnamefont {A.}~\bibnamefont {Al\`{u}}},\
	}\href {\doibase 10.1126/science.aar7709} {\bibfield  {journal} {\bibinfo
			{journal} {Science}\ }\textbf {\bibinfo {volume} {363}},\ \bibinfo {pages}
		{eaar7709} (\bibinfo {year} {2019})}\BibitemShut {NoStop}%
	\bibitem [{\citenamefont {Yoon}\ \emph {et~al.}(2018)\citenamefont {Yoon},
		\citenamefont {Choi}, \citenamefont {Hahn}, \citenamefont {Kim},
		\citenamefont {Song}, \citenamefont {Yang}, \citenamefont {Lee},
		\citenamefont {Kim}, \citenamefont {Lee}, \citenamefont {Shin}, \citenamefont
		{Lee},\ and\ \citenamefont {Berini}}]{RN3.1.6}%
	\BibitemOpen
	\bibfield  {author} {\bibinfo {author} {\bibfnamefont {J.~W.}\ \bibnamefont
			{Yoon}}, \bibinfo {author} {\bibfnamefont {Y.}~\bibnamefont {Choi}}, \bibinfo
		{author} {\bibfnamefont {C.}~\bibnamefont {Hahn}}, \bibinfo {author}
		{\bibfnamefont {G.}~\bibnamefont {Kim}}, \bibinfo {author} {\bibfnamefont
			{S.~H.}\ \bibnamefont {Song}}, \bibinfo {author} {\bibfnamefont {K.-Y.}\
			\bibnamefont {Yang}}, \bibinfo {author} {\bibfnamefont {J.~Y.}\ \bibnamefont
			{Lee}}, \bibinfo {author} {\bibfnamefont {Y.}~\bibnamefont {Kim}}, \bibinfo
		{author} {\bibfnamefont {C.~S.}\ \bibnamefont {Lee}}, \bibinfo {author}
		{\bibfnamefont {J.~K.}\ \bibnamefont {Shin}}, \bibinfo {author}
		{\bibfnamefont {H.-S.}\ \bibnamefont {Lee}}, \ and\ \bibinfo {author}
		{\bibfnamefont {P.}~\bibnamefont {Berini}},\ }\href {\doibase
		10.1038/s41586-018-0523-2} {\bibfield  {journal} {\bibinfo  {journal}
			{Nature}\ }\textbf {\bibinfo {volume} {562}},\ \bibinfo {pages} {86}
		(\bibinfo {year} {2018})}\BibitemShut {NoStop}%
	\bibitem [{\citenamefont {Zhong}\ \emph {et~al.}(2020)\citenamefont {Zhong},
		\citenamefont {Ozdemir}, \citenamefont {Eisfeld}, \citenamefont {Metelmann},\
		and\ \citenamefont {El-Ganainy}}]{RN3.1.7}%
	\BibitemOpen
	\bibfield  {author} {\bibinfo {author} {\bibfnamefont {Q.}~\bibnamefont
			{Zhong}}, \bibinfo {author} {\bibfnamefont {S.}~\bibnamefont {Ozdemir}},
		\bibinfo {author} {\bibfnamefont {A.}~\bibnamefont {Eisfeld}}, \bibinfo
		{author} {\bibfnamefont {A.}~\bibnamefont {Metelmann}}, \ and\ \bibinfo
		{author} {\bibfnamefont {R.}~\bibnamefont {El-Ganainy}},\ }\href {\doibase
		10.1103/PhysRevApplied.13.014070} {\bibfield  {journal} {\bibinfo  {journal}
			{Phys. Rev. Applied}\ }\textbf {\bibinfo {volume} {13}},\ \bibinfo {pages}
		{014070} (\bibinfo {year} {2020})}\BibitemShut {NoStop}%
	\bibitem [{\citenamefont {Heiss}(2016)}]{RN1.8.1}%
	\BibitemOpen
	\bibfield  {author} {\bibinfo {author} {\bibfnamefont {D.}~\bibnamefont
			{Heiss}},\ }\href {\doibase 10.1038/nphys3864} {\bibfield  {journal}
		{\bibinfo  {journal} {Nature Physics}\ }\textbf {\bibinfo {volume} {12}},\
		\bibinfo {pages} {823} (\bibinfo {year} {2016})}\BibitemShut {NoStop}%
	\bibitem [{\citenamefont {Zhang}\ \emph {et~al.}(2019)\citenamefont {Zhang},
		\citenamefont {Jiang},\ and\ \citenamefont {Chan}}]{RN1.9.1}%
	\BibitemOpen
	\bibfield  {author} {\bibinfo {author} {\bibfnamefont {X.-L.}\ \bibnamefont
			{Zhang}}, \bibinfo {author} {\bibfnamefont {T.}~\bibnamefont {Jiang}}, \ and\
		\bibinfo {author} {\bibfnamefont {C.~T.}\ \bibnamefont {Chan}},\ }\href
	{\doibase 10.1038/s41377-019-0200-8} {\bibfield  {journal} {\bibinfo
			{journal} {Light: Science \& Applications}\ }\textbf {\bibinfo {volume} {8}},\
		\bibinfo {pages} {88} (\bibinfo {year} {2019})}\BibitemShut {NoStop}%
	\bibitem [{\citenamefont {Hassan}\ \emph {et~al.}(2017)\citenamefont {Hassan},
		\citenamefont {Zhen}, \citenamefont {Solja\ifmmode \check{c}\else
			\v{c}\fi{}i\ifmmode~\acute{c}\else \'{c}\fi{}}, \citenamefont {Khajavikhan},\
		and\ \citenamefont {Christodoulides}}]{RN1.9.2}%
	\BibitemOpen
	\bibfield  {author} {\bibinfo {author} {\bibfnamefont {A.~U.}\ \bibnamefont
			{Hassan}}, \bibinfo {author} {\bibfnamefont {B.}~\bibnamefont {Zhen}},
		\bibinfo {author} {\bibfnamefont {M.}~\bibnamefont {Solja\ifmmode
				\check{c}\else \v{c}\fi{}i\ifmmode~\acute{c}\else \'{c}\fi{}}}, \bibinfo
		{author} {\bibfnamefont {M.}~\bibnamefont {Khajavikhan}}, \ and\ \bibinfo
		{author} {\bibfnamefont {D.~N.}\ \bibnamefont {Christodoulides}},\ }\href
	{\doibase 10.1103/PhysRevLett.118.093002} {\bibfield  {journal} {\bibinfo
			{journal} {Phys. Rev. Lett.}\ }\textbf {\bibinfo {volume} {118}},\ \bibinfo
		{pages} {093002} (\bibinfo {year} {2017})}\BibitemShut {NoStop}%
	\bibitem [{\citenamefont {Liu}\ \emph {et~al.}(2020)\citenamefont {Liu},
		\citenamefont {Li}, \citenamefont {Wang}, \citenamefont {Ke}, \citenamefont
		{Qin}, \citenamefont {Wang}, \citenamefont {Liu}, \citenamefont {Gao},
		\citenamefont {Berini},\ and\ \citenamefont {Lu}}]{RN1.10}%
	\BibitemOpen
	\bibfield  {author} {\bibinfo {author} {\bibfnamefont {Q.}~\bibnamefont
			{Liu}}, \bibinfo {author} {\bibfnamefont {S.}~\bibnamefont {Li}}, \bibinfo
		{author} {\bibfnamefont {B.}~\bibnamefont {Wang}}, \bibinfo {author}
		{\bibfnamefont {S.}~\bibnamefont {Ke}}, \bibinfo {author} {\bibfnamefont
			{C.}~\bibnamefont {Qin}}, \bibinfo {author} {\bibfnamefont {K.}~\bibnamefont
			{Wang}}, \bibinfo {author} {\bibfnamefont {W.}~\bibnamefont {Liu}}, \bibinfo
		{author} {\bibfnamefont {D.}~\bibnamefont {Gao}}, \bibinfo {author}
		{\bibfnamefont {P.}~\bibnamefont {Berini}}, \ and\ \bibinfo {author}
		{\bibfnamefont {P.}~\bibnamefont {Lu}},\ }\href {\doibase
		10.1103/PhysRevLett.124.153903} {\bibfield  {journal} {\bibinfo  {journal}
			{Phys. Rev. Lett.}\ }\textbf {\bibinfo {volume} {124}},\ \bibinfo {pages}
		{153903} (\bibinfo {year} {2020})}\BibitemShut {NoStop}%
	\bibitem [{\citenamefont {Ke}\ \emph {et~al.}(2016)\citenamefont {Ke},
		\citenamefont {Wang}, \citenamefont {Qin}, \citenamefont {Long},
		\citenamefont {Wang},\ and\ \citenamefont {Lu}}]{RN1.11}%
	\BibitemOpen
	\bibfield  {author} {\bibinfo {author} {\bibfnamefont {S.}~\bibnamefont
			{Ke}}, \bibinfo {author} {\bibfnamefont {B.}~\bibnamefont {Wang}}, \bibinfo
		{author} {\bibfnamefont {C.}~\bibnamefont {Qin}}, \bibinfo {author}
		{\bibfnamefont {H.}~\bibnamefont {Long}}, \bibinfo {author} {\bibfnamefont
			{K.}~\bibnamefont {Wang}}, \ and\ \bibinfo {author} {\bibfnamefont
			{P.}~\bibnamefont {Lu}},\ }\href
	{http://opg.optica.org/jlt/abstract.cfm?URI=jlt-34-22-5258} {\bibfield
		{journal} {\bibinfo  {journal} {Journal of Lightwave Technology}\ }\textbf
		{\bibinfo {volume} {34}},\ \bibinfo {pages} {5258} (\bibinfo {year}
		{2016})}\BibitemShut {NoStop}%
	\bibitem [{\citenamefont {Song}\ \emph {et~al.}(2021)\citenamefont {Song},
		\citenamefont {Odeh}, \citenamefont {Z\'{u}\~{n}iga-P\'{e}rez}, \citenamefont
		{Kant\'{e}},\ and\ \citenamefont {Genevet}}]{RN1.11.2}%
	\BibitemOpen
	\bibfield  {author} {\bibinfo {author} {\bibfnamefont {Q.}~\bibnamefont
			{Song}}, \bibinfo {author} {\bibfnamefont {M.}~\bibnamefont {Odeh}}, \bibinfo
		{author} {\bibfnamefont {J.}~\bibnamefont {Z\'{u}\~{n}iga-P\'{e}re}}, \bibinfo
		{author} {\bibfnamefont {B.}~\bibnamefont {Kant\'{e}}}, \ and\ \bibinfo {author}
		{\bibfnamefont {P.}~\bibnamefont {Genevet}},\ }\href {\doibase
		10.1126/science.abj3179} {\bibfield  {journal} {\bibinfo  {journal}
			{Science}\ }\textbf {\bibinfo {volume} {373}},\ \bibinfo {pages} {1133}
		(\bibinfo {year} {2021})}\BibitemShut {NoStop}%
	\bibitem [{\citenamefont {Vahala}(2003)}]{RN3.1.8}%
	\BibitemOpen
	\bibfield  {author} {\bibinfo {author} {\bibfnamefont {K.~J.}\ \bibnamefont
			{Vahala}},\ }\href {\doibase 10.1038/nature01939} {\bibfield  {journal}
		{\bibinfo  {journal} {Nature}\ }\textbf {\bibinfo {volume} {424}},\ \bibinfo
		{pages} {839} (\bibinfo {year} {2003})}\BibitemShut {NoStop}%
	\bibitem [{\citenamefont {Indukuri}\ \emph {et~al.}(2019)\citenamefont
		{Indukuri}, \citenamefont {Bar-David}, \citenamefont {Mazurski},\ and\
		\citenamefont {Levy}}]{RN1.12.2}%
	\BibitemOpen
	\bibfield  {author} {\bibinfo {author} {\bibfnamefont {S.~R. K.~C.}\
			\bibnamefont {Indukuri}}, \bibinfo {author} {\bibfnamefont {J.}~\bibnamefont
			{Bar-David}}, \bibinfo {author} {\bibfnamefont {N.}~\bibnamefont {Mazurski}},
		\ and\ \bibinfo {author} {\bibfnamefont {U.}~\bibnamefont {Levy}},\ }\href
	{\doibase 10.1021/acsnano.9b05730} {\bibfield  {journal} {\bibinfo  {journal}
			{ACS Nano}\ }\textbf {\bibinfo {volume} {13}},\ \bibinfo {pages} {11770}
		(\bibinfo {year} {2019})}\BibitemShut {NoStop}%
	\bibitem [{\citenamefont {Li}\ \emph {et~al.}(2020)\citenamefont {Li},
		\citenamefont {Dong}, \citenamefont {Wang}, \citenamefont {Cheng},
		\citenamefont {Ho}, \citenamefont {Zhang}, \citenamefont {Wen}, \citenamefont
		{Zhang}, \citenamefont {Chan}, \citenamefont {Al\`u}, \citenamefont {Qiu},\
		and\ \citenamefont {Chen}}]{RN1.13.2}%
	\BibitemOpen
	\bibfield  {author} {\bibinfo {author} {\bibfnamefont {A.}~\bibnamefont
			{Li}}, \bibinfo {author} {\bibfnamefont {J.}~\bibnamefont {Dong}}, \bibinfo
		{author} {\bibfnamefont {J.}~\bibnamefont {Wang}}, \bibinfo {author}
		{\bibfnamefont {Z.}~\bibnamefont {Cheng}}, \bibinfo {author} {\bibfnamefont
			{J.~S.}\ \bibnamefont {Ho}}, \bibinfo {author} {\bibfnamefont
			{D.}~\bibnamefont {Zhang}}, \bibinfo {author} {\bibfnamefont
			{J.}~\bibnamefont {Wen}}, \bibinfo {author} {\bibfnamefont {X.-L.}\
			\bibnamefont {Zhang}}, \bibinfo {author} {\bibfnamefont {C.~T.}\ \bibnamefont
			{Chan}}, \bibinfo {author} {\bibfnamefont {A.}~\bibnamefont {Al\`u}},
		\bibinfo {author} {\bibfnamefont {C.-W.}\ \bibnamefont {Qiu}}, \ and\
		\bibinfo {author} {\bibfnamefont {L.}~\bibnamefont {Chen}},\ }\href {\doibase
		10.1103/PhysRevLett.125.187403} {\bibfield  {journal} {\bibinfo  {journal}
			{Phys. Rev. Lett.}\ }\textbf {\bibinfo {volume} {125}},\ \bibinfo {pages}
		{187403} (\bibinfo {year} {2020})}\BibitemShut {NoStop}%
	\bibitem [{\citenamefont {Heiss}\ and\ \citenamefont {Harney}(2001)}]{RN1.14}%
	\BibitemOpen
	\bibfield  {author} {\bibinfo {author} {\bibfnamefont {W.~D.}\ \bibnamefont
			{Heiss}}\ and\ \bibinfo {author} {\bibfnamefont {H.~L.}\ \bibnamefont
			{Harney}},\ }\href {\doibase 10.1007/s100530170017} {\bibfield  {journal}
		{\bibinfo  {journal} {The European Physical Journal D - Atomic, Molecular,
				Optical and Plasma Physics}\ }\textbf {\bibinfo {volume} {17}},\ \bibinfo
		{pages} {149} (\bibinfo {year} {2001})}\BibitemShut {NoStop}%
	\bibitem [{\citenamefont {Peng}\ \emph
		{et~al.}(2014{\natexlab{b}})\citenamefont {Peng}, \citenamefont {\"{O}zdemir},
		\citenamefont {Chen}, \citenamefont {Nori},\ and\ \citenamefont
		{Yang}}]{RN2.2.1}%
	\BibitemOpen
	\bibfield  {author} {\bibinfo {author} {\bibfnamefont {B.}~\bibnamefont
			{Peng}}, \bibinfo {author} {\bibfnamefont {\c{S}~K.}\ \bibnamefont {\"{O}zdemir}},
		\bibinfo {author} {\bibfnamefont {W.}~\bibnamefont {Chen}}, \bibinfo {author}
		{\bibfnamefont {F.}~\bibnamefont {Nori}}, \ and\ \bibinfo {author}
		{\bibfnamefont {L.}~\bibnamefont {Yang}},\ }\href {\doibase
		10.1038/ncomms6082} {\bibfield  {journal} {\bibinfo  {journal} {Nature
				Communications}\ }\textbf {\bibinfo {volume} {5}},\ \bibinfo {pages} {5082}
		(\bibinfo {year} {2014}{\natexlab{b}})}\BibitemShut {NoStop}%
	\bibitem [{\citenamefont {Li}\ \emph {et~al.}(2015)\citenamefont {Li},
		\citenamefont {Yu},\ and\ \citenamefont {Wu}}]{RN2.2.2}%
	\BibitemOpen
	\bibfield  {author} {\bibinfo {author} {\bibfnamefont {J.}~\bibnamefont
			{Li}}, \bibinfo {author} {\bibfnamefont {R.}~\bibnamefont {Yu}}, \ and\
		\bibinfo {author} {\bibfnamefont {Y.}~\bibnamefont {Wu}},\ }\href {\doibase
		10.1103/PhysRevA.92.053837} {\bibfield  {journal} {\bibinfo  {journal} {Phys.
				Rev. A}\ }\textbf {\bibinfo {volume} {92}},\ \bibinfo {pages} {053837}
		(\bibinfo {year} {2015})}\BibitemShut {NoStop}%
	\bibitem [{\citenamefont {Aspelmeyer}\ \emph {et~al.}(2014)\citenamefont
		{Aspelmeyer}, \citenamefont {Kippenberg},\ and\ \citenamefont
		{Marquardt}}]{RN2.3.1}%
	\BibitemOpen
	\bibfield  {author} {\bibinfo {author} {\bibfnamefont {M.}~\bibnamefont
			{Aspelmeyer}}, \bibinfo {author} {\bibfnamefont {T.~J.}\ \bibnamefont
			{Kippenberg}}, \ and\ \bibinfo {author} {\bibfnamefont {F.}~\bibnamefont
			{Marquardt}},\ }\href {\doibase 10.1103/RevModPhys.86.1391} {\bibfield
		{journal} {\bibinfo  {journal} {Rev. Mod. Phys.}\ }\textbf {\bibinfo {volume}
			{86}},\ \bibinfo {pages} {1391} (\bibinfo {year} {2014})}\BibitemShut
	{NoStop}%
	\bibitem [{\citenamefont {Wang}\ and\ \citenamefont {Clerk}(2012)}]{RN2.6.1}%
	\BibitemOpen
	\bibfield  {author} {\bibinfo {author} {\bibfnamefont {Y.-D.}\ \bibnamefont
			{Wang}}\ and\ \bibinfo {author} {\bibfnamefont {A.~A.}\ \bibnamefont
			{Clerk}},\ }\href {\doibase 10.1103/PhysRevLett.108.153603} {\bibfield
		{journal} {\bibinfo  {journal} {Phys. Rev. Lett.}\ }\textbf {\bibinfo
			{volume} {108}},\ \bibinfo {pages} {153603} (\bibinfo {year}
		{2012})}\BibitemShut {NoStop}%
	\bibitem [{\citenamefont {Tian}(2012)}]{RN2.6.2}%
	\BibitemOpen
	\bibfield  {author} {\bibinfo {author} {\bibfnamefont {L.}~\bibnamefont
			{Tian}},\ }\href {\doibase 10.1103/PhysRevLett.108.153604} {\bibfield
		{journal} {\bibinfo  {journal} {Phys. Rev. Lett.}\ }\textbf {\bibinfo
			{volume} {108}},\ \bibinfo {pages} {153604} (\bibinfo {year}
		{2012})}\BibitemShut {NoStop}%
	\bibitem [{\citenamefont {Liu}\ \emph {et~al.}(2013)\citenamefont {Liu},
		\citenamefont {Xiao}, \citenamefont {Luan},\ and\ \citenamefont
		{Wong}}]{RN3.1.11}%
	\BibitemOpen
	\bibfield  {author} {\bibinfo {author} {\bibfnamefont {Y.-C.}\ \bibnamefont
			{Liu}}, \bibinfo {author} {\bibfnamefont {Y.-F.}\ \bibnamefont {Xiao}},
		\bibinfo {author} {\bibfnamefont {X.}~\bibnamefont {Luan}}, \ and\ \bibinfo
		{author} {\bibfnamefont {C.~W.}\ \bibnamefont {Wong}},\ }\href {\doibase
		10.1103/PhysRevLett.110.153606} {\bibfield  {journal} {\bibinfo  {journal}
			{Phys. Rev. Lett.}\ }\textbf {\bibinfo {volume} {110}},\ \bibinfo {pages}
		{153606} (\bibinfo {year} {2013})}\BibitemShut {NoStop}%
	\bibitem [{\citenamefont {Liao}\ and\ \citenamefont {Nori}(2013)}]{RN3.1.13}%
	\BibitemOpen
	\bibfield  {author} {\bibinfo {author} {\bibfnamefont {J.-Q.}\ \bibnamefont
			{Liao}}\ and\ \bibinfo {author} {\bibfnamefont {F.}~\bibnamefont {Nori}},\
	}\href {\doibase 10.1103/PhysRevA.88.023853} {\bibfield  {journal} {\bibinfo
			{journal} {Phys. Rev. A}\ }\textbf {\bibinfo {volume} {88}},\ \bibinfo
		{pages} {023853} (\bibinfo {year} {2013})}\BibitemShut {NoStop}%
	\bibitem [{\citenamefont {Yin}\ \emph {et~al.}(2013)\citenamefont {Yin},
		\citenamefont {Li}, \citenamefont {Zhang},\ and\ \citenamefont
		{Duan}}]{RN3.1.14}%
	\BibitemOpen
	\bibfield  {author} {\bibinfo {author} {\bibfnamefont {Z.-q.}\ \bibnamefont
			{Yin}}, \bibinfo {author} {\bibfnamefont {T.}~\bibnamefont {Li}}, \bibinfo
		{author} {\bibfnamefont {X.}~\bibnamefont {Zhang}}, \ and\ \bibinfo {author}
		{\bibfnamefont {L.~M.}\ \bibnamefont {Duan}},\ }\href {\doibase
		10.1103/PhysRevA.88.033614} {\bibfield  {journal} {\bibinfo  {journal} {Phys.
				Rev. A}\ }\textbf {\bibinfo {volume} {88}},\ \bibinfo {pages} {033614}
		(\bibinfo {year} {2013})}\BibitemShut {NoStop}%
	\bibitem [{\citenamefont {L\"u}\ \emph {et~al.}(2015)\citenamefont {L\"u},
		\citenamefont {Jing}, \citenamefont {Ma},\ and\ \citenamefont
		{Wu}}]{RN3.1.12}%
	\BibitemOpen
	\bibfield  {author} {\bibinfo {author} {\bibfnamefont {X.-Y.}\ \bibnamefont
			{L\"u}}, \bibinfo {author} {\bibfnamefont {H.}~\bibnamefont {Jing}}, \bibinfo
		{author} {\bibfnamefont {J.-Y.}\ \bibnamefont {Ma}}, \ and\ \bibinfo {author}
		{\bibfnamefont {Y.}~\bibnamefont {Wu}},\ }\href {\doibase
		10.1103/PhysRevLett.114.253601} {\bibfield  {journal} {\bibinfo  {journal}
			{Phys. Rev. Lett.}\ }\textbf {\bibinfo {volume} {114}},\ \bibinfo {pages}
		{253601} (\bibinfo {year} {2015})}\BibitemShut {NoStop}%
	\bibitem [{\citenamefont {Cao}\ \emph {et~al.}(2016)\citenamefont {Cao},
		\citenamefont {Mi}, \citenamefont {Gao}, \citenamefont {He}, \citenamefont
		{Yang}, \citenamefont {Wang}, \citenamefont {Zhang},\ and\ \citenamefont
		{Wang}}]{RN3.1.15}%
	\BibitemOpen
	\bibfield  {author} {\bibinfo {author} {\bibfnamefont {C.}~\bibnamefont
			{Cao}}, \bibinfo {author} {\bibfnamefont {S.-C.}\ \bibnamefont {Mi}},
		\bibinfo {author} {\bibfnamefont {Y.-P.}\ \bibnamefont {Gao}}, \bibinfo
		{author} {\bibfnamefont {L.-Y.}\ \bibnamefont {He}}, \bibinfo {author}
		{\bibfnamefont {D.}~\bibnamefont {Yang}}, \bibinfo {author} {\bibfnamefont
			{T.-J.}\ \bibnamefont {Wang}}, \bibinfo {author} {\bibfnamefont
			{R.}~\bibnamefont {Zhang}}, \ and\ \bibinfo {author} {\bibfnamefont
			{C.}~\bibnamefont {Wang}},\ }\href {\doibase 10.1038/srep22920} {\bibfield
		{journal} {\bibinfo  {journal} {Scientific Reports}\ }\textbf {\bibinfo
			{volume} {6}},\ \bibinfo {pages} {22920} (\bibinfo {year}
		{2016})}\BibitemShut {NoStop}%
	\bibitem [{\citenamefont {Shen}\ \emph {et~al.}(2018)\citenamefont {Shen},
		\citenamefont {Zhang}, \citenamefont {Chen}, \citenamefont {Sun},
		\citenamefont {Zou}, \citenamefont {Guo}, \citenamefont {Zou},\ and\
		\citenamefont {Dong}}]{RN2.6.3}%
	\BibitemOpen
	\bibfield  {author} {\bibinfo {author} {\bibfnamefont {Z.}~\bibnamefont
			{Shen}}, \bibinfo {author} {\bibfnamefont {Y.-L.}\ \bibnamefont {Zhang}},
		\bibinfo {author} {\bibfnamefont {Y.}~\bibnamefont {Chen}}, \bibinfo {author}
		{\bibfnamefont {F.-W.}\ \bibnamefont {Sun}}, \bibinfo {author} {\bibfnamefont
			{X.-B.}\ \bibnamefont {Zou}}, \bibinfo {author} {\bibfnamefont {G.-C.}\
			\bibnamefont {Guo}}, \bibinfo {author} {\bibfnamefont {C.-L.}\ \bibnamefont
			{Zou}}, \ and\ \bibinfo {author} {\bibfnamefont {C.-H.}\ \bibnamefont
			{Dong}},\ }\href {\doibase 10.1038/s41467-018-04187-8} {\bibfield  {journal}
		{\bibinfo  {journal} {Nature Communications}\ }\textbf {\bibinfo {volume}
			{9}},\ \bibinfo {pages} {1797} (\bibinfo {year} {2018})}\BibitemShut
	{NoStop}%
	\bibitem [{\citenamefont {Mao}\ \emph {et~al.}(2022)\citenamefont {Mao},
		\citenamefont {Qin}, \citenamefont {Yang}, \citenamefont {Wang},
		\citenamefont {Wang}, \citenamefont {Li}, \citenamefont {Xue},\ and\
		\citenamefont {Long}}]{RN3.1.4}%
	\BibitemOpen
	\bibfield  {author} {\bibinfo {author} {\bibfnamefont {X.}~\bibnamefont
			{Mao}}, \bibinfo {author} {\bibfnamefont {G.-Q.}\ \bibnamefont {Qin}},
		\bibinfo {author} {\bibfnamefont {H.}~\bibnamefont {Yang}}, \bibinfo {author}
		{\bibfnamefont {Z.}~\bibnamefont {Wang}}, \bibinfo {author} {\bibfnamefont
			{M.}~\bibnamefont {Wang}}, \bibinfo {author} {\bibfnamefont {G.-Q.}\
			\bibnamefont {Li}}, \bibinfo {author} {\bibfnamefont {P.}~\bibnamefont
			{Xue}}, \ and\ \bibinfo {author} {\bibfnamefont {G.-L.}\ \bibnamefont
			{Long}},\ }\href {\doibase 10.1103/PhysRevA.105.033526} {\bibfield  {journal}
		{\bibinfo  {journal} {Phys. Rev. A}\ }\textbf {\bibinfo {volume} {105}},\
		\bibinfo {pages} {033526} (\bibinfo {year} {2022})}\BibitemShut {NoStop}%
	\bibitem [{\citenamefont {Xiao}\ \emph {et~al.}(2021)\citenamefont {Xiao},
		\citenamefont {Xiong}, \citenamefont {Zhao},\ and\ \citenamefont
		{Yin}}]{RN3.1.16}%
	\BibitemOpen
	\bibfield  {author} {\bibinfo {author} {\bibfnamefont {K.-W.}\ \bibnamefont
			{Xiao}}, \bibinfo {author} {\bibfnamefont {A.}~\bibnamefont {Xiong}},
		\bibinfo {author} {\bibfnamefont {N.}~\bibnamefont {Zhao}}, \ and\ \bibinfo
		{author} {\bibfnamefont {Z.-q.}\ \bibnamefont {Yin}},\ }\href {\doibase
		https://doi.org/10.1002/que2.62} {\bibfield  {journal} {\bibinfo  {journal}
			{Quantum Engineering}\ }\textbf {\bibinfo {volume} {3}},\ \bibinfo {pages}
		{e62} (\bibinfo {year} {2021})}\BibitemShut {NoStop}%
	\bibitem [{\citenamefont {Dong}\ \emph {et~al.}(2012)\citenamefont {Dong},
		\citenamefont {Fiore}, \citenamefont {Kuzyk},\ and\ \citenamefont
		{Wang}}]{RN3.1.9}%
	\BibitemOpen
	\bibfield  {author} {\bibinfo {author} {\bibfnamefont {C.}~\bibnamefont
			{Dong}}, \bibinfo {author} {\bibfnamefont {V.}~\bibnamefont {Fiore}},
		\bibinfo {author} {\bibfnamefont {M.~C.}\ \bibnamefont {Kuzyk}}, \ and\
		\bibinfo {author} {\bibfnamefont {H.}~\bibnamefont {Wang}},\ }\href {\doibase
		10.1126/science.1228370} {\bibfield  {journal} {\bibinfo  {journal}
			{Science}\ }\textbf {\bibinfo {volume} {338}},\ \bibinfo {pages} {1609}
		(\bibinfo {year} {2012})}\BibitemShut {NoStop}%
	\bibitem [{\citenamefont {Li}\ \emph {et~al.}(2013)\citenamefont {Li},
		\citenamefont {Ren}, \citenamefont {Liu},\ and\ \citenamefont
		{Xiao}}]{RN2.5.1}%
	\BibitemOpen
	\bibfield  {author} {\bibinfo {author} {\bibfnamefont {H.-K.}\ \bibnamefont
			{Li}}, \bibinfo {author} {\bibfnamefont {X.-X.}\ \bibnamefont {Ren}},
		\bibinfo {author} {\bibfnamefont {Y.-C.}\ \bibnamefont {Liu}}, \ and\
		\bibinfo {author} {\bibfnamefont {Y.-F.}\ \bibnamefont {Xiao}},\ }\href
	{\doibase 10.1103/PhysRevA.88.053850} {\bibfield  {journal} {\bibinfo
			{journal} {Phys. Rev. A}\ }\textbf {\bibinfo {volume} {88}},\ \bibinfo
		{pages} {053850} (\bibinfo {year} {2013})}\BibitemShut {NoStop}%
	\bibitem [{\citenamefont {Zhang}\ \emph
		{et~al.}(2019{\natexlab{b}})\citenamefont {Zhang}, \citenamefont {Song},
		\citenamefont {Ai}, \citenamefont {Wang}, \citenamefont {Yang},\ and\
		\citenamefont {Deng}}]{RN3.1.10}%
	\BibitemOpen
	\bibfield  {author} {\bibinfo {author} {\bibfnamefont {H.}~\bibnamefont
			{Zhang}}, \bibinfo {author} {\bibfnamefont {X.-K.}\ \bibnamefont {Song}},
		\bibinfo {author} {\bibfnamefont {Q.}~\bibnamefont {Ai}}, \bibinfo {author}
		{\bibfnamefont {H.}~\bibnamefont {Wang}}, \bibinfo {author} {\bibfnamefont
			{G.-J.}\ \bibnamefont {Yang}}, \ and\ \bibinfo {author} {\bibfnamefont
			{F.-G.}\ \bibnamefont {Deng}},\ }\href {\doibase 10.1364/OE.27.007384}
	{\bibfield  {journal} {\bibinfo  {journal} {Optics Express}\ }\textbf
		{\bibinfo {volume} {27}},\ \bibinfo {pages} {7384} (\bibinfo {year}
		{2019}{\natexlab{b}})}\BibitemShut {NoStop}%
	\bibitem [{\citenamefont {Milburn}\ \emph {et~al.}(2015)\citenamefont
		{Milburn}, \citenamefont {Doppler}, \citenamefont {Holmes}, \citenamefont
		{Portolan}, \citenamefont {Rotter},\ and\ \citenamefont {Rabl}}]{RN3.1.1}%
	\BibitemOpen
	\bibfield  {author} {\bibinfo {author} {\bibfnamefont {T.~J.}\ \bibnamefont
			{Milburn}}, \bibinfo {author} {\bibfnamefont {J.}~\bibnamefont {Doppler}},
		\bibinfo {author} {\bibfnamefont {C.~A.}\ \bibnamefont {Holmes}}, \bibinfo
		{author} {\bibfnamefont {S.}~\bibnamefont {Portolan}}, \bibinfo {author}
		{\bibfnamefont {S.}~\bibnamefont {Rotter}}, \ and\ \bibinfo {author}
		{\bibfnamefont {P.}~\bibnamefont {Rabl}},\ }\href {\doibase
		10.1103/PhysRevA.92.052124} {\bibfield  {journal} {\bibinfo  {journal} {Phys.
				Rev. A}\ }\textbf {\bibinfo {volume} {92}},\ \bibinfo {pages} {052124}
		(\bibinfo {year} {2015})}\BibitemShut {NoStop}%
	\bibitem [{\citenamefont {Yin}\ \emph {et~al.}(2020)\citenamefont {Yin},
		\citenamefont {Luo}, \citenamefont {Zhang}, \citenamefont {Lin},
		\citenamefont {Tian}, \citenamefont {Li}, \citenamefont {Wang}, \citenamefont
		{Duan}, \citenamefont {Huang},\ and\ \citenamefont {Du}}]{RN3.1.2}%
	\BibitemOpen
	\bibfield  {author} {\bibinfo {author} {\bibfnamefont {P.}~\bibnamefont
			{Yin}}, \bibinfo {author} {\bibfnamefont {X.}~\bibnamefont {Luo}}, \bibinfo
		{author} {\bibfnamefont {L.}~\bibnamefont {Zhang}}, \bibinfo {author}
		{\bibfnamefont {S.}~\bibnamefont {Lin}}, \bibinfo {author} {\bibfnamefont
			{T.}~\bibnamefont {Tian}}, \bibinfo {author} {\bibfnamefont {R.}~\bibnamefont
			{Li}}, \bibinfo {author} {\bibfnamefont {Z.}~\bibnamefont {Wang}}, \bibinfo
		{author} {\bibfnamefont {C.}~\bibnamefont {Duan}}, \bibinfo {author}
		{\bibfnamefont {P.}~\bibnamefont {Huang}}, \ and\ \bibinfo {author}
		{\bibfnamefont {J.}~\bibnamefont {Du}},\ }\href {\doibase
		10.1088/0256-307x/37/10/100301} {\bibfield  {journal} {\bibinfo  {journal}
			{Chinese Physics Letters}\ }\textbf {\bibinfo {volume} {37}},\ \bibinfo
		{pages} {100301} (\bibinfo {year} {2020})}\BibitemShut {NoStop}%
	\bibitem [{\citenamefont {Zhang}\ \emph
		{et~al.}(2018{\natexlab{b}})\citenamefont {Zhang}, \citenamefont {Wang},
		\citenamefont {Hou},\ and\ \citenamefont {Chan}}]{RN3.1.3}%
	\BibitemOpen
	\bibfield  {author} {\bibinfo {author} {\bibfnamefont {X.-L.}\ \bibnamefont
			{Zhang}}, \bibinfo {author} {\bibfnamefont {S.}~\bibnamefont {Wang}},
		\bibinfo {author} {\bibfnamefont {B.}~\bibnamefont {Hou}}, \ and\ \bibinfo
		{author} {\bibfnamefont {C.~T.}\ \bibnamefont {Chan}},\ }\href {\doibase
		10.1103/PhysRevX.8.021066} {\bibfield  {journal} {\bibinfo  {journal} {Phys.
				Rev. X}\ }\textbf {\bibinfo {volume} {8}},\ \bibinfo {pages} {021066}
		(\bibinfo {year} {2018}{\natexlab{b}})}\BibitemShut {NoStop}%
\end{thebibliography}

\end{document}